\documentclass[11pt]{article}

\usepackage[utf8]{inputenc}
\usepackage[T1]{fontenc}
\usepackage{amsmath,amssymb,amsfonts}
\usepackage{graphicx}
\usepackage{booktabs}
\usepackage{hyperref}
\usepackage{natbib}
\usepackage{geometry}
\usepackage{caption}
\usepackage{subcaption}
\usepackage{enumitem}
\usepackage{bm}
\usepackage{float}

\hypersetup{
    colorlinks=true,
    linkcolor=blue,
    citecolor=blue,
    urlcolor=blue
}

\title{Kelly Betting as Bayesian Model Evaluation:\\ A Framework for Time-Updating Probabilistic Forecasts}
\author{Michael Beuoy\\
inpredictable.com\\
\texttt{michael.beuoy@gmail.com}}
\date{}

\begin{document}

\maketitle

\begin{abstract}
This paper proposes a new way of evaluating the accuracy and validity of probabilistic forecasts that change over time (such as an in-game win probability model, or an election forecast). Under this approach, each model to be evaluated is treated as a canonical Kelly bettor, and the models are pitted against each other in an iterative betting contest. The growth or decline of each model's bankroll serves as the evaluation metric. Under this approach, market consensus probabilities and implied model credibilities can be updated real time as each model updates, and do not require one to wait for the final outcome. Using a simulation model, it will be shown that this method is in general more accurate than traditional average log-loss and Brier score methods at distinguishing a correct model from an incorrect model. This Kelly approach is shown to have a direct mathematical and conceptual analogue to Bayesian inference, with bankroll serving as a proxy for Bayesian credibility.
\end{abstract}

%----------------------------------------------------------------------
\section{Introduction}
\label{sec:intro}
%----------------------------------------------------------------------

Forecast evaluation can be a fraught topic, particularly with the general public. Probabilistic thinking does not come naturally to most, making it difficult to understand the performance of a prediction that conveys something short of ironclad certainty. Typically, one will often see a probabilistic forecast collapsed into a binary proposition and evaluated as such. For example, all probabilities greater than 50\% (e.g.\ 51\%, 90\%, 99.9999\%) are viewed equally as a prediction of certainty for the outcome in question. Conversely, you will sometimes see probabilistic forecasts viewed as practically unfalsifiable---as long as you predicted a probability greater than 0\% (even if it was 0.1\%), your forecast can never be wrong.

The waters become even muddier when evaluating forecasts that change over time. Outside of sporting events, the most widely known time-updating forecasts are those that attempt to forecast the outcome of elections, such as the models and probabilities once published at FiveThirtyEight. Despite updating on a daily basis, election forecasts are often judged by the accuracy of their last forecast, published the morning of the election. On its surface, this seems somewhat odd. What value is a forecast where you will likely know the outcome within 24 hours anyway? Surely the forecasts produced 3, 4, 5 months in advance of the election would be more valuable and a better test of forecasting skill?

Returning to sports, the most common time-updating forecasts are in-game win probability models. And these are often judged by their ``worst'' prediction, particularly in the case of improbable comebacks. When the New England Patriots overcame a 28--3 deficit in Super Bowl LI, the infinitesimal win probability given to the Patriots at their lowest point was held up by some as proof of the uselessness of win probability as a metric.

Fortunately, there do exist statistically sound tools for general forecast evaluation, such as log loss, Brier score, and calibration tests. These tests do a good job of punishing overconfident predictions while rewarding well-calibrated predictions.

However, these evaluation tools are not specifically suited to time-updating forecasts. For example, if Model A estimates the probability of an outcome initially at 70\%, but then updates it to 30\% at a later time, how would we compare that to Model B, which estimated 30\% all along? Assume in this case that the outcome did occur. A traditional approach would just be to take the average log loss or Brier score of each estimate and then compare. But such an approach ignores the order in which each model makes its predictions. Couldn't one argue that Model B performed better because it got to the 30\% answer quicker than the first model? Even though it ends up with a worse average log loss score? And should we view Model A with less credibility because of that? Or was Model B just lucky, having hit upon the right answer for the wrong reasons, and Model A was the one appropriately reacting to new information as it came to light?

These questions are difficult to answer in general, but what this paper proposes is a method of forecast evaluation that provides an unambiguous, accurate, and fundamentally Bayesian way of answering this question. We'll start with a simple example involving some friendly wagers.

%----------------------------------------------------------------------
\section{A Simple Example}
\label{sec:simple_example}
%----------------------------------------------------------------------

Suppose we have two modelers, Bob and Alice, who both have their own independent forecasts for an upcoming basketball game. In addition, both Bob and Alice can update their forecast after the completion of each quarter.

Bob and Alice's estimates are summarized in Table~\ref{tab:bob_alice_probs} below.

\begin{table}[H]
\centering
\caption{Bob and Alice's home team win probability estimates at the beginning of each quarter}
\label{tab:bob_alice_probs}
\begin{tabular}{lcc}
\toprule
& \multicolumn{2}{c}{\textbf{home team win probability}} \\
\cmidrule(lr){2-3}
\textbf{forecast as of:} & Bob & Alice \\
\midrule
Start of 1st Quarter & 80\% & 50\% \\
Start of 2nd Quarter & 50\% & 50\% \\
Start of 3rd Quarter & 50\% & 80\% \\
Start of 4th Quarter & 80\% & 80\% \\
\bottomrule
\end{tabular}
\end{table}

Let's say that the home team ended up winning this game. How would we evaluate the accuracy of Bob and Alice's forecasts? A common approach in this situation would be to calculate a Brier score or log loss for each forecast made, and then average that number. Under this evaluation, Bob and Alice performed equally well, with Brier scores of 0.145 and log loss of 0.66 (using a base 2 log).

What if, instead of comparing Brier scores or log loss, Bob and Alice were to wager against each other real time, using their respective probability models? It turns out that if we assume both Bob and Alice follow the Kelly criterion in determining their bet sizes, we can calculate a unique odds at each point of the game in which Bob and Alice would agree to a wager.

For example, if we assume Bob and Alice both start with a \$50 bankroll, they would agree to a bet at 7/13 odds, or an implied probability of 65\%. Bob would be willing to make a bet of \$21.43 at those odds, and Alice would be willing to accept a bet (as the bookmaker) of \$21.43 at those same odds.

As the game progresses and Bob and Alice update their probabilities, they can continue to find a mutually agreed upon bet size and odds that reflects both their latest probability estimates, as well as the existing wagers they have already made. For example, after the 1st quarter, when Bob has updated his probability estimate to 50\%, while Alice remains at 50\%, they would then agree to a bet of \$16.48 at even odds. However, in this case, Bob has become the bookmaker and is accepting \$16.48 from Alice, the bettor.

It turns out that under this approach, we end up with a clear winner between Bob and Alice. At the end of the game, Bob has lost \$9.45 of his \$50 by betting with Alice, who has now added that \$9.45 to her \$50 bankroll.

The math to calculate these bets is straightforward, and will be demonstrated in subsequent sections.

%----------------------------------------------------------------------
\section{Binary Probabilities with Multiple Bettors}
\label{sec:binary}
%----------------------------------------------------------------------

We will start with the simple example of binary probabilities, and then extend the method to an arbitrary number of mutually exclusive outcomes.

\subsection{Optimal Kelly betting with existing bets}
\label{sec:kelly_existing}

The Kelly criterion is a bet-sizing strategy that prescribes the percentage of bankroll one should wager on an outcome, given the odds offered and the estimated true probability of the outcome. The most familiar version of the Kelly criterion expresses the fraction of bankroll to be wagered, given the estimated probability and odds:
\[
f = p - \frac{1 - p}{o}
\]
Where $f$ is the fraction of bankroll to wager, $p$ is the estimated probability of the outcome to occur, and $o$ is the odds offered.

This formula can be generalized to the situation where a bettor has existing bets on the outcome. Denote $w$ as a bettor's \emph{win shares}, and define it as the money a bettor would stand to win (or \emph{lose}, if $w$ is negative) if the outcome in question occurred.

In this situation, the Kelly Criterion can be generalized as (see Appendix~\ref{app:kelly_existing} for a derivation):
\begin{equation}
f = p - \frac{(1-p)}{o}\left(1+\frac{w}{b}\right)
\label{eq:kelly_existing}
\end{equation}

This reduces to the more familiar version of the Kelly criterion when a bettor has no existing win shares ($w=0$).

\subsection{Determining a ``market clearing price'' with multiple Kelly bettors and existing bets}
\label{sec:market_clearing}

Using formula~\eqref{eq:kelly_existing}, we can now calculate the desired bet size for a bettor that reflects both their estimate of the outcome probability $p$ as well as any bets they have already made, as reflected by their win shares amount, $w$.

For the case of $n$ bettors, denote $p_i$ as the outcome probability assigned by the $i$th bettor, $b_i$ as the bankroll held by the $i$th bettor and $w_i$ as the win shares held by the $i$th bettor. Assume that each bettor follows formula~\eqref{eq:kelly_existing} when determining what fraction of their bankroll to invest in a proposed set of odds. One can establish a formula for the market clearing price where, in aggregate, all bets are matched up with willing bookmakers among the $n$ bettors.

Formula~\eqref{eq:market_prob} below is the market clearing price, expressed in terms of the implied probability from the market odds (refer to Appendix~\ref{app:market_clearing} for the derivation):
\begin{equation}
\text{market probability} = \frac{\sum p_i b_i}{1-\sum p_i w_i}
\label{eq:market_prob}
\end{equation}

This is a generalization of a formula from \citet{beygelzimer2012}, which found that in the case of a market without existing bets, the ``market clearing'' price is the bankroll-weighted probability estimate of the market participants. In the general case where there are existing bets to account for, the market probability contains a denominator term which is 1 minus the sum product of each bettor's win shares times each bettor's probability. Formula~\eqref{eq:market_prob} reduces to the original 2012 result if all win shares are set to zero.

Using formulas~\eqref{eq:kelly_existing} and~\eqref{eq:market_prob}, we can evaluate multiple forecast models against each other in a de facto betting contest. A key advantage of this approach is that it allows for a real time updating of the performance of each model. In other words, you do not need to wait until the bet is ``settled'' in order to update evaluation of model performance. Using the market consensus probability, $m_p$, each model's portfolio of bets can be marked to market.

Returning to our Bob and Alice example, when Bob reduced his original probability estimate from 80\% to 50\% (and Alice stayed put at 50\%), the marked-to-market value of Bob's portfolio dropped from \$50 to \$45. If we view bankroll as analogous to credibility (in a Bayesian sense), our Bayesian prior that Bob and Alice were equally probable to have the correct model, has now been updated to Bob being 45\% likely to have the correct model, and Alice upgraded to 55\%.

\subsection{Procedure for evaluating multiple forecast models against each other in the case of binary probabilities}
\label{sec:procedure}

We now have a straightforward algorithm for evaluating real-time forecast models against each other. The steps are as follows:

\begin{itemize}
    \item \textbf{Step 1:} Calculate the market clearing probability according to formula~\eqref{eq:market_prob}:
    \[
    \text{market probability} = \frac{\sum p_i b_i}{1-\sum p_i w_i}
    \]
    This is a simple calculation based solely on each bettor's bankroll, outstanding win shares, and latest probability estimate.

    \item \textbf{Step 2:} Convert the market probability into odds via the formula $o = (1-p)/p$ and then calculate each bettor's wager as a fraction of their bankroll according to formula~\eqref{eq:kelly_existing}:
    \[
    f = p - \frac{(1-p)}{o}\left(1+\frac{w}{b}\right)
    \]

    \item \textbf{Step 3} (optional): Calculate each bettor's ``marked to market'' bankroll. The market consensus probability can be used to evaluate each bettor's expected bankroll, which is equal to their bankroll plus the expected value of their win shares. This provides a running evaluation of how each bettor is doing in this prediction contest. One does not have to wait for the event to be over and the bets to be settled.

    \item \textbf{Step 4:} Update each bettor's bankroll and win shares based on the fraction of bankroll bet from step 2. Win shares are updated in accordance with the market clearing odds. For example, if a bettor is wagering 20\% of their \$100 bankroll at 2-1 market odds, you reduce their bankroll by \$20 and increase their win shares by \$60 (= [bet amount] $\times$ [odds + 1])

    \item \textbf{Step 5:} Repeat Steps 1--4 as new information is revealed and the bettors update their probability estimates.

    \item \textbf{Step 6:} Settling the bet. In the event of a win, each bettor's bankroll is increased/decreased by their outstanding win shares amount. Note that, by necessity, some bettors will have negative win shares and will thus see a reduction in their bankroll. They are the bookmakers paying out on winning wagers. In the event of a loss, all win shares have a value of zero and each bettor is left with their existing bankroll as their end result.
\end{itemize}

Steps 1--6 show how one can create an evaluation metric for competing real-time forecast models. One can find a similar approach in \citet{sethi2025}, in which the authors applied a Kelly-based betting approach to evaluate the accuracy of models predicting the outcome of the 2024 US presidential and congressional elections. However, in that approach, the betting odds are determined by Polymarket, a crypto-based prediction market, rather than set by consensus as in this paper's approach.

Returning to the example from the introduction, here is how these steps would play out:

\begin{table}[H]
\centering
\caption{Illustration of the methodology with a hypothetical example where two modelers wager against each other on a basketball game.}
\label{tab:bob_alice_full}
\resizebox{\textwidth}{!}{%
\begin{tabular}{l cc cc cc c cc cc cc}
\toprule
& \multicolumn{2}{c}{\textbf{win prob}} & \multicolumn{2}{c}{\textbf{bankroll}} & \multicolumn{2}{c}{\textbf{win shares}} & \textbf{market} & \multicolumn{2}{c}{\textbf{credibility}} & \multicolumn{2}{c}{\textbf{bet amount}} & \multicolumn{2}{c}{\textbf{+/- win shares}} \\
\cmidrule(lr){2-3}\cmidrule(lr){4-5}\cmidrule(lr){6-7}\cmidrule(lr){9-10}\cmidrule(lr){11-12}\cmidrule(lr){13-14}
\textbf{qtr} & Bob & Alice & Bob & Alice & Bob & Alice & \textbf{prob} & Bob & Alice & Bob & Alice & Bob & Alice \\
\midrule
Q1 & 80\% & 50\% & 0.50 & 0.50 & 0.00 & 0.00 & 65\% & 50\% & 50\% & 0.21 & $-$0.21 & 0.33 & $-$0.33 \\
Q2 & 50\% & 50\% & 0.29 & 0.71 & 0.33 & $-$0.33 & 50\% & 45\% & 55\% & $-$0.16 & 0.16 & $-$0.33 & 0.33 \\
Q3 & 50\% & 80\% & 0.45 & 0.55 & 0.00 & 0.00 & 66\% & 45\% & 55\% & $-$0.22 & 0.22 & $-$0.33 & 0.33 \\
Q4 & 80\% & 80\% & 0.67 & 0.33 & $-$0.33 & 0.33 & 80\% & 41\% & 59\% & 0.27 & $-$0.27 & 0.33 & $-$0.33 \\
\midrule
\textbf{Final} & & & \textbf{0.41} & \textbf{0.59} & \textbf{0.00} & \textbf{0.00} & & & & & & & \\
\bottomrule
\end{tabular}%
}
\end{table}

Bob, by continually being one step behind Alice, sees 9\% of his credibility siphoned away over the course of the game. Also note that in this case, the actual outcome of the game is immaterial to the final result for Bob and Alice. In general, it can be shown that when all market participants agree on the probabilities at any point in the game, all bets get effectively settled, and every market participant's win shares go to zero.

Compared to traditional methods like average log loss or Brier score, this metric has a more tangible, real world interpretation: if you let each model bet according to the Kelly criterion, models can be evaluated by how their bankroll shrinks or grows as a result of this betting contest. In the following section we will show that bankroll in this context has both a conceptual and mathematical analogue to Bayesian credibility. To demonstrate this, we will first show how the algorithm above can be generalized to the case of multinomial probabilities.

%----------------------------------------------------------------------
\section{The General Case with Multinomial Probabilities and Multiple Bettors}
\label{sec:multinomial}
%----------------------------------------------------------------------

The approach described in Section~\ref{sec:binary} will now be generalized to account for an arbitrary number of mutually exclusive outcomes. Real world examples from the sporting world include in-match soccer probabilities (win/loss/draw) or in-season team futures, such as predicting the winner of an MLB division.

\subsection{Kelly betting with multiple outcomes and existing bets}
\label{sec:kelly_multi}

To best model multinomial outcomes, we will dispense with the bankroll/bet distinction of the previous section, and instead specify the positions held by each model in terms of the bankroll they would ultimately be left with in the event of each outcome. This avoids having to arbitrarily set aside one outcome as the ``no bet'' option. This approach also aligns with that taken by Kelly in his original paper \citep{kelly1956}, and will result in a generalization of his original result.

Define $w_i$ as the win shares held on the $i$th outcome by the bettor. In this situation there is no additional bankroll to speak of, so $w_i$ represents the sum total of funds the bettor would be left with in the event of the $i$th outcome. Under this definition, a bettor with no existing bets to speak of would simply have a uniform $w_i$ for each outcome, where $w_i$ is just their bankroll.

To illustrate with the Bob and Alice example from the introduction, after the 1st quarter, rather than saying that Bob has a bankroll of \$21.53 and win shares of \$32.97, we instead describe Bob's position in terms of his ending bankroll under each potential outcome:

\begin{itemize}
    \item $w_{\text{home\_win}}$ = \$61.54 (=\$21.53 + \$32.97)
    \item $w_{\text{home\_loss}}$ = \$21.53 (=\$21.53 + \$0.00)
\end{itemize}

We now define $p_i$ as the probability assigned to the $i$th outcome by the bettor, and $m_i$ as the corresponding market probability, meaning the probability implied by the available market odds for each outcome.

We can now calculate how a bettor's win shares shift as the result of a new round of Kelly wagering. We will denote the bettor's updated win shares as $w'_i$, and can be expressed as (see Appendix~\ref{app:kelly_multi} for a derivation):
\begin{equation}
w'_i = \frac{p_i}{m_i} \sum_{i=1}^{n} m_i w_i
\label{eq:kelly_multi}
\end{equation}

This is a generalization of a result from John Kelly's original paper \citep{kelly1956}, which found that when the market odds are fair, the optimal wagering strategy is to bet in proportion to one's own probability estimates for each outcome, \emph{independent of the market odds being offered.}

To see how this generalizes Kelly's original result, note that betting a fraction $p_i$ of your bankroll $b$ at market odds $o_i$ means that you will end up with the following bankroll under each potential outcome $i$:
\[
w_i = \frac{p_i}{m_i} b
\]
Where $m_i$ is the probability of outcome $i$ implied by the market odds $o_i$ ($m_i = 1/(1+o_i)$).

This is a less general version of formula~\eqref{eq:kelly_multi}, specific to the situation where a bettor has no existing bets. The generalization in formula~\eqref{eq:kelly_multi} states that when you have existing bets, you are in effect betting in proportion to your marked to market bankroll, which is the expected value of your existing positions, according to the market probabilities:
\[
\text{``marked to market'' bankroll} = \sum_{i=1}^{n} m_i w_i
\]

\subsubsection{Calculating market clearing odds with multiple outcomes and multiple bettors}
\label{sec:market_multi}

In the case of $n$ possible outcomes and $m$ Kelly bettors, we will now calculate the implied market clearing odds as a function of each bettor's existing positions and current probability estimates. We define the following:

\begin{itemize}[nosep]
    \item $w_{ij}$ = win shares held on the $i$th outcome by the $j$th bettor
    \item $p_{ij}$ = probability estimate of the $i$th outcome from the $j$th bettor
    \item $o_i$ = the ``market clearing'' odds for the $i$th outcome. These are the odds in which all bettors have matching ``bookmakers'' for the wagers they wish to place
    \item $m_i$ = the probability of the $i$th outcome implied by the market odds $o_i$
    \item $\Delta_{ij}$ = win shares to be purchased (or sold) on the $i$th outcome by the $j$th bettor
\end{itemize}

If we consider $p_{ij}$ and $w_{kj}$ as two $n \times m$ matrices $\bm{p}$ and $\bm{w}$, and the market probabilities $m_i$ as a vector $\bm{m}$, the condition for market-clearing odds can be written as (see Appendix~\ref{app:market_multi} for a derivation):
\begin{equation}
(\bm{p}\bm{w}^T) \bm{m} = \bm{m}
\label{eq:eigenvector}
\end{equation}

Written this way, we see that the market clearing probability $\bm{m}$ is an eigenvector of the matrix $\bm{p}\bm{w}^T$ with eigenvalue 1. We will refer to the matrix $\bm{p}\bm{w}^T$ as $\bm{M}$. $\bm{M}$ is an $n \times n$ matrix in which the $ij$th element is the sum product of the probability estimates for outcome $i$ times the win shares held on outcome $j$, summed over all market participants. The matrix $\bm{M}$ can be interpreted as a conditional probability matrix of sorts, where the $ij$th element tells you the posterior probability of outcome $j$, in the event that outcome $i$ has occurred. Or, in conditional probability notation, $M_{ij} = P(j \mid i)$.

\subsubsection{A self-evaluating market}
\label{sec:self_eval}

If we multiply our matrix equation~\eqref{eq:eigenvector} by the matrix $\bm{w}^T$ on both sides we get the following:

\[
\bm{w}^T(\bm{p}\bm{w}^T) \bm{m} = \bm{w}^T\bm{m}
\]

Or, regrouping terms:
\begin{equation}
(\bm{w}^T\bm{p})(\bm{w}^T\bm{m}) = \bm{w}^T\bm{m}
\label{eq:self_eval}
\end{equation}

The vector $\bm{w}^T\bm{m}$ is a vector of length $m$, in which the $i$th element is the sum product of the market probabilities times the $i$th bettor's portfolio, (recall that $\bm{w}$ tells you the bankroll each bettor will end up with for each potential outcome). Put another way, this vector is the expected value of each bettor's portfolio, using the market consensus probabilities to calculate the expected value. If we define this vector as $\bm{c}$, where $c$ stands for credibility, we can see from equation~\eqref{eq:self_eval} that $\bm{c}$ is an eigenvector of the matrix $\bm{w}^T\bm{p}$.

The matrix $\bm{w}^T\bm{p}$ is an $m \times m$ matrix with an intuitive interpretation. The $ij$th element of this matrix is the sum product of the $i$th bettor's probability estimates times the $j$th bettor's portfolio. In other words, this matrix tells us what each bettor thinks of every other bettor's portfolio (including their own). We will define the matrix $\bm{w}^T\bm{p}$ as $\bm{S}$ (where $S$ stands for self-evaluating market). We can now rewrite equation~\eqref{eq:self_eval} more simply as:

\[
\bm{S}\bm{c} = \bm{c}
\]

Our credibility vector $\bm{c}$ is an eigenvector of $\bm{S}$, with an eigenvalue of 1. The matrix $\bm{S}$ is also a left-stochastic matrix and thus the eigenvector $\bm{c}$ with eigenvalue of 1 is guaranteed to exist.

The credibility vector $\bm{c}$ and the market clearing probability vector $\bm{m}$ are connected via the following matrix equations:

\[
\bm{c} = \bm{w}^T\bm{m}
\]
\[
\bm{m} = \bm{p}\bm{c}
\]

So, one need only solve for either the eigenvector of $\bm{p}\bm{w}^T$ or $\bm{w}^T\bm{p}$ and then use one of the equations above to convert $\bm{m}$ to $\bm{c}$ or $\bm{c}$ to $\bm{m}$.

The procedure described in Section~\ref{sec:procedure} can now be extended to the general case of multinomial probabilities. In Step 1, formula~\eqref{eq:market_prob} for the market probability is replaced with the eigenvector of the market odds matrix $\bm{M}$. Step 2 is replaced by formula~\eqref{eq:delta} from the Appendix, which tells you how to adjust each bettor's win shares, given a set of market odds $m$.

Several illustrations of the approach across multiple sports and types of forecasts are provided in Appendix~\ref{app:examples} of this paper.

\subsubsection{Bankroll as credibility: Connection to Bayes' Theorem}
\label{sec:bayes}

Formula~\eqref{eq:kelly_multi} has a direct connection to Bayes' theorem. To show this, instead of treating bankroll as a financial metric, we will interpret it instead as a likelihood of a model being the correct model. This was the motivation for having the total of all bankrolls sum to 1. How can we reframe the Kelly approach in Bayesian terms?

The hypothesis we wish to evaluate (in a Bayesian way) is the likelihood of each model being correct. The evidence we use to evaluate the hypothesis and update our priors are the actual outcomes that each model is trying to predict. Using the standard version of Bayes' formula, define $P(A|B)$ as the likelihood that model $A$ is the correct model, given that outcome $B$ occurred. Then, according to Bayes' theorem, $P(A|B)$ can be calculated as:
\[
P(A|B) = \frac{P(B|A) \cdot P(A)}{P(B)}
\]

We can now show that, under a suitable interpretation of our Kelly-based betting contest, formula~\eqref{eq:kelly_multi} is equivalent to Bayes' theorem above.

If we interpret bankroll as a likelihood, then the term $P(A|B)$ is just $w_B$, the bettor's ending bankroll conditioned upon the event $B$ having occurred.

Moving to the right side of Bayes' formula, $P(B|A)$ is the probability that $B$ will occur, given that $A$ is the correct model. Using our previous definitions, this is just $p_B$, the probability assigned to outcome $B$ by model $A$.

$P(B)$ is the total probability that outcome $B$ will occur, taking into account all models. Trusting the wisdom of the crowd, $P(B)$ is the probability of $B$ implied by the market odds, or $m_B$.

$P(A)$ is our prior estimate of the likelihood that model $A$ is the correct model. We define this with the help of the market probabilities as well. We do this by calculating the expected bankroll (i.e.\ likelihood) of bettor $A$'s portfolio, using the market probabilities to assess expected value:
\[
\sum_{i=1}^{n} m_i w_i
\]

To summarize:
\begin{itemize}
    \item $P(A|B) \to w_B$ (the probability that $A$ is the correct model given outcome $B$)
    \item $P(B|A) \to p_B$ (the probability of $B$ if model $A$ is the correct model)
    \item $P(B) \to m_B$ (the market consensus probability of $B$)
    \item $P(A) \to \sum_{i=1}^{n} m_i w_i$ (our Bayesian prior, the likelihood that $A$ is the correct model)
\end{itemize}

So, plugging these terms into Bayes' theorem gives us:
\[
w_B = \frac{p_B}{m_B} \sum_{i=1}^{n} m_i w_i
\]

Which is identical to formula~\eqref{eq:kelly_multi}, with different subscripts.

%----------------------------------------------------------------------
\section{Simulating Various Models to Demonstrate the Value of This Approach}
\label{sec:simulations}
%----------------------------------------------------------------------

\subsection{Creating a hypothetical sports contest to model and then simulating}
\label{sec:sim_setup}

To illustrate the value of this approach in identifying more accurate models, we will model and simulate a volleyball-like contest in which the actual probability of a team winning any given point is constant. The rules of this simplified game are that the first team to 100 points wins, but they must win by at least 2 points.

The true win probabilities at any given game state can be calculated straightforwardly, given the probability of winning a single point. In what follows, we will create one ``correct'' model, and various ``incorrect'' models to then demonstrate how this paper's approach fares when attempting to identify the correct model from the actual outcome of simulated games. In particular, we will see when this approach is superior to more traditional approaches of forecast evaluation, such as log loss and Brier score. These incorrect models were crafted in an attempt to recreate common situations in which a forecast model can go wrong.

\subsection{When the incorrect model has the wrong point probability}
\label{sec:sim_wrong_prob}

For the first example, we will assume that the actual probability of Team A winning a point against team B is 50\%. We will then pit two models against each other in an in-game prediction contest:

\begin{itemize}
    \item Correct model---Assumes Team A has a 50\% probability to win each point
    \item Incorrect model---Assumes Team A has a 53\% probability to win each point
\end{itemize}

As an example, if the score was 10--15 in favor of Team B, the correct model would calculate the probability of Team A winning the game to be 35.2\%. But the incorrect modeler would calculate that same probability as 66.2\%.

Below are the results of simulating 10,000 games, using a simulated point win probability of 50\%. We will use three evaluation metrics: ending bankroll (this paper's approach), average log loss score (straight average across all points played), and average Brier score (straight average across all points played).

For each simulation round, we record which model received the higher score according to these three metrics. Here are the results. ``Accuracy'' is defined as the percent of times that the correct model scored the higher metric.

\begin{table}[H]
\centering
\caption{Distinguishing between a correct model with 0.50 point probability and an incorrect model with 0.53 point probability}
\label{tab:sim_50v53}
\begin{tabular}{lc}
\toprule
\textbf{Evaluation Metric} & \textbf{Accuracy} \\
\midrule
Kelly Bankroll & 55.1\% \\
Average Log Loss & 49.9\% \\
Average Brier Score & 49.9\% \\
\bottomrule
\end{tabular}
\end{table}

So, 55.1\% of the time, the correct model ends the game with a higher bankroll than the incorrect model. But 49.9\% of the time, the correct model ended with a better log loss score than the incorrect model.

Sticking with this example, we can also simulate when the correct probability is 0.53:

\begin{itemize}
    \item Correct model---Assumes Team A has a 53\% probability to win each point
    \item Incorrect model---Assumes Team A has a 50\% probability to win each point
\end{itemize}

Simulating 10,000 times, we get the following result:

\begin{table}[H]
\centering
\caption{Distinguishing between a correct model with 0.53 point probability and an incorrect model with 0.50 point probability}
\label{tab:sim_53v50}
\begin{tabular}{lc}
\toprule
\textbf{Evaluation Metric} & \textbf{Accuracy} \\
\midrule
Kelly Bankroll & 76.3\% \\
Average Log Loss & 80.5\% \\
Average Brier Score & 80.5\% \\
\bottomrule
\end{tabular}
\end{table}

Contrary to the first example, the Kelly approach was less accurate in identifying the correct model, compared to averaging of either log loss or Brier scores. However, we will see in a subsequent section that when the contest is iterated, the Kelly approach outperforms log loss and Brier methods over the long term across a broad range of point probability scenarios.

\subsection{When the incorrect model has faulty recency bias}
\label{sec:sim_recency}

For the second example, the incorrect model will start with the correct point probability, 0.50. However, as the game progresses, the incorrect model begins updating its point probabilities based on the actual point outcomes. More specifically, the incorrect model treats the prior 10 points played with 10\% credibility when projecting the future probability of winning a point:

\begin{itemize}
    \item Incorrect model Team A point probability = (90\% $\times$ 0.50) + (10\% $\times$ [Percent of points won by Team A for prior 10 points played])
\end{itemize}

Below are the results of simulating 10,000 games in this manner, and recording which approach was more accurate in identifying the correct model:

\begin{table}[H]
\centering
\caption{Distinguishing between a correct model with 0.50 point probability and an incorrect model with faulty recency bias}
\label{tab:sim_recency}
\begin{tabular}{lc}
\toprule
\textbf{Evaluation Metric} & \textbf{Accuracy} \\
\midrule
Kelly Bankroll & 96.0\% \\
Average Log Loss & 73.1\% \\
Average Brier Score & 80.2\% \\
\bottomrule
\end{tabular}
\end{table}

\subsection{When the incorrect model uses variables with no predictive value}
\label{sec:sim_random}

In this example, we will simulate a situation when an incorrect model starts with the correct point probability of 0.50, but then updates that probability as the game progresses based on a variable that has no predictive value. The point probability the incorrect model uses follows a random walk, in which after each point the forecasted point probability is adjusted up or down by the following amount:

\begin{itemize}
    \item ([random number between 0 and 1] $-$ 0.5) / 35
\end{itemize}

The modeled point probabilities are restricted to stay within the values of 0.40 and 0.60. These parameters were chosen in order to create a plausibly incorrect model, rather than an outlandishly incorrect model.

Below are the results of simulating 10,000 games in this manner, and recording which approach was more accurate in identifying the correct model.

\begin{table}[H]
\centering
\caption{Distinguishing between a correct model with 0.50 point probability and an incorrect model that uses a variable with no predictive value}
\label{tab:sim_random}
\begin{tabular}{lc}
\toprule
\textbf{Evaluation Metric} & \textbf{Accuracy} \\
\midrule
Kelly Bankroll & 74.4\% \\
Average Log Loss & 57.6\% \\
Average Brier Score & 58.3\% \\
\bottomrule
\end{tabular}
\end{table}

As seen from the results in Tables~\ref{tab:sim_recency} and~\ref{tab:sim_random}, the Kelly approach is particularly adept at identifying incorrect models that are unnecessarily volatile.

%----------------------------------------------------------------------
\section{Adding Iteration to the Simulations}
\label{sec:iteration}
%----------------------------------------------------------------------

\subsection{Utilizing the Bayesian Nature of the Kelly Approach}
\label{sec:iteration_bayes}

A key advantage of the Kelly approach is its Bayesian nature, allowing it to have ``memory'' when dealing with an iterated comparison of models. More specifically, suppose you have a sequence of games in which you have two models forecasting the in-game probability of winning. As we would in any typical betting contest, we can let each model's ending bankroll from the prior game be used as the starting bankroll for the new game. As shown in Section~\ref{sec:bayes}, bankroll in this context has a conceptual and mathematical analog to Bayesian credibility. Each in-game betting result informs our priors of how we start the subsequent game.

When iterated, it will be shown in the next section that the Kelly approach is superior to traditional log loss and Brier methods across a robust set of scenarios.

\subsection{Simulating an Iterated Betting Contest}
\label{sec:iteration_sim}

For this simulation, we will extend our simulation in Section~\ref{sec:sim_wrong_prob} to a $11 \times 11$ grid of ``correct'' and ``incorrect'' models, in which both models will take on all distinct combinations of the following point probabilities: 0.45, 0.46, 0.47, 0.48, 0.49, 0.50, 0.51, 0.52, 0.53, 0.54, and 0.55. The simulation will then iterate over 50 games in sequence. For the Kelly approach, the ending bankroll from one game carries over to the start of the next game. For the log loss and Brier score approaches, models will be evaluated by average log loss and Brier score over all points played to that point.

This 50 game approach is simulated 1,000 times for each point probability combination. The accuracy after 1, 10, 25, and 50 contests is shown below.

Here are the results after a single game of betting. The Kelly approach ``wins'' if it more frequently identifies the correct model than if one would just compare average Log Loss or Brier Score for each model. A Scenario here means a specific choice of ``correct'' model point probability (e.g.\ 0.54) and ``incorrect'' model point probability (e.g.\ 0.47).

\begin{table}[H]
\centering
\caption{Model evaluation accuracy after a single game}
\label{tab:iter_1game}
\begin{tabular}{lcc}
\toprule
\textbf{Method} & \multicolumn{2}{c}{\textbf{Scenarios Won}} \\
\cmidrule(lr){2-3}
& $n$ & \% \\
\midrule
Kelly & 50 & 45\% \\
Tie & 14 & 13\% \\
Log Loss or Brier & 46 & 42\% \\
\midrule
Total & 110 & 100\% \\
\bottomrule
\end{tabular}
\end{table}

As we saw in Section~\ref{sec:sim_wrong_prob}, after a single round, there are scenarios where Log Loss and Brier score are more accurate. The Kelly approach holds just a slight advantage in terms of Scenarios ``won''.

Here are the results after just 5 rounds. ``5 rounds'' in this case means that the contest was repeated 5 times, with each model's Kelly bankroll at the end of a round carrying over to the start of the next round.

\begin{table}[H]
\centering
\caption{Model evaluation accuracy after 5 games}
\label{tab:iter_5games}
\begin{tabular}{lcc}
\toprule
\textbf{Method} & \multicolumn{2}{c}{\textbf{Scenarios Won}} \\
\cmidrule(lr){2-3}
& $n$ & \% \\
\midrule
Kelly & 98 & 89\% \\
Tie & 1 & 1\% \\
Log Loss or Brier & 11 & 10\% \\
\midrule
Total & 110 & 100\% \\
\bottomrule
\end{tabular}
\end{table}

Even after 5 rounds of betting, the advantage of the Kelly approach is apparent. Of the 110 scenarios, it was most accurate in 98 of those, and tied for best accuracy in 1 scenario. It only lost to either average log loss or Brier score 11 times (10\%).

Here are the results after 25 rounds of betting. There are now even fewer scenarios (3 out of 110), in which the Kelly approach doesn't at least tie for best accuracy.

\begin{table}[H]
\centering
\caption{Model evaluation accuracy after 25 games}
\label{tab:iter_25games}
\begin{tabular}{lcc}
\toprule
\textbf{Method} & \multicolumn{2}{c}{\textbf{Scenarios Won}} \\
\cmidrule(lr){2-3}
& $n$ & \% \\
\midrule
Kelly & 76 & 69\% \\
Tie & 31 & 28\% \\
Log Loss or Brier & 3 & 3\% \\
\midrule
Total & 110 & 100\% \\
\bottomrule
\end{tabular}
\end{table}

And finally, here are the results after 50 rounds of betting. Only 2 scenarios (out of 110) remain in which the Kelly approach did not at least tie for best accuracy.

\begin{table}[H]
\centering
\caption{Model evaluation accuracy after 50 games}
\label{tab:iter_50games}
\begin{tabular}{lcc}
\toprule
\textbf{Method} & \multicolumn{2}{c}{\textbf{Scenarios Won}} \\
\cmidrule(lr){2-3}
& $n$ & \% \\
\midrule
Kelly & 61 & 55\% \\
Tie & 47 & 43\% \\
Log Loss or Brier & 2 & 2\% \\
\midrule
Total & 110 & 100\% \\
\bottomrule
\end{tabular}
\end{table}

Refer to the charts in Appendix Section~\ref{app:sim_exhibits} for additional detail on which specific scenarios the Kelly model over or under performed against Log Loss and Brier Score.

%----------------------------------------------------------------------
\section{In-Game Evaluations of Model Credibility}
\label{sec:ingame}
%----------------------------------------------------------------------

\subsection{Evaluating models prior to completion of the contest}
\label{sec:ingame_intro}

Another advantage of the Kelly approach is that it allows for ``real time'' evaluation of model accuracy and credibility, prior to the completion of the contest. In the preceding sections, the comparison to average log loss and Brier score methods required knowing the outcome of the game before calculating a log loss or Brier score. Using the simulation examples from the previous sections, we will show how the Kelly approach can identify which model is more accurate (in a Bayesian way) in the early stages of the game.

In this paper, we refer to a model's ``credibility'' and a model's ``bankroll'' interchangeably.

\subsection{In-game evaluations where the incorrect model has the wrong point probability}
\label{sec:ingame_wrong}

The table below shows the average implied credibility each model has after 10, 25, 50, and 100 points played. This is based on the same simulation of 10,000 games from the previous section.

\begin{table}[H]
\centering
\caption{Correct model point probability: 0.50, Incorrect model point probability: 0.53}
\label{tab:ingame_50v53}
\begin{tabular}{lcc}
\toprule
& \multicolumn{2}{c}{\textbf{Average Model Credibility}} \\
\cmidrule(lr){2-3}
\textbf{After point:} & \textbf{correct model} & \textbf{incorrect model} \\
\midrule
10 & 50.2\% & 49.8\% \\
25 & 50.6\% & 49.4\% \\
50 & 51.1\% & 48.9\% \\
100 & 52.1\% & 47.9\% \\
\bottomrule
\end{tabular}
\end{table}

\begin{table}[H]
\centering
\caption{Correct model point probability: 0.53, Incorrect model point probability: 0.50}
\label{tab:ingame_53v50}
\begin{tabular}{lcc}
\toprule
& \multicolumn{2}{c}{\textbf{Average Model Credibility}} \\
\cmidrule(lr){2-3}
\textbf{After point:} & \textbf{correct model} & \textbf{incorrect model} \\
\midrule
10 & 50.3\% & 49.7\% \\
25 & 50.7\% & 49.3\% \\
50 & 51.3\% & 48.7\% \\
100 & 52.5\% & 47.5\% \\
\bottomrule
\end{tabular}
\end{table}

While not showing sizable gains in credibility (there is a ceiling to what one can statistically conclude from just a sample of points), these results do show that the Kelly approach is able to tilt its credibility estimates towards the correct model early on in the contest, and prior to knowing the outcome.

\subsection{In-game evaluations where the incorrect model has faulty recency bias}
\label{sec:ingame_recency}

The table below shows the same results for the situation in which the correct point probability is 0.50, but the incorrect model erroneously adjusts its point probabilities in response to recent in-game performance.

\begin{table}[H]
\centering
\caption{Correct model point probability of 0.50 versus a faulty recency bias model}
\label{tab:ingame_recency}
\begin{tabular}{lcc}
\toprule
& \multicolumn{2}{c}{\textbf{Average Model Credibility}} \\
\cmidrule(lr){2-3}
\textbf{Point} & \textbf{correct model} & \textbf{incorrect model} \\
\midrule
10 & 50.1\% & 49.9\% \\
25 & 52.1\% & 47.9\% \\
50 & 54.7\% & 45.3\% \\
100 & 58.2\% & 41.8\% \\
\bottomrule
\end{tabular}
\end{table}

\subsection{In-game evaluations where the incorrect model uses variables with no predictive value}
\label{sec:ingame_random}

Here are the results where the incorrect model point probabilities drift according to a random variable.

\begin{table}[H]
\centering
\caption{Correct model point probability of 0.50 versus a model using non-predictive variables}
\label{tab:ingame_random}
\begin{tabular}{lcc}
\toprule
& \multicolumn{2}{c}{\textbf{Average Model Credibility}} \\
\cmidrule(lr){2-3}
\textbf{Point} & \textbf{correct model} & \textbf{incorrect model} \\
\midrule
10 & 50.6\% & 49.4\% \\
25 & 51.9\% & 48.1\% \\
50 & 54.1\% & 45.9\% \\
100 & 57.9\% & 42.1\% \\
\bottomrule
\end{tabular}
\end{table}

As we see from this example and the prior recency bias one, the Kelly approach is particularly good at downgrading the credibility of models that show unnecessary volatility.

%----------------------------------------------------------------------
\section{Advantages of the Methodology}
\label{sec:advantages}
%----------------------------------------------------------------------

This approach has multiple advantages over traditional scoring methods such as log loss and Brier score, including:

\begin{itemize}
    \item Using simulations, the preceding sections showed that the Kelly approach is more accurate in separating correct models from incorrect models across a broad range of scenarios

    \item The evaluation metric is time order-dependent, and, in general, rewards models that are more prescient and less volatile than their competitors, as we saw with the Bob and Alice example

    \item Average log loss and Brier score when applied to in-game forecasts can sometimes be skewed by arbitrary over-weighting of predictions, such as if the model probabilities are changing by very small amounts, each updated probability gets weighted equally in the log loss or Brier score. Under this paper's approach, small changes to probabilities lead to commensurately small bets, and thus do not have a material impact on the evaluation metric

    \item Model credibility can be evaluated real time as the event progresses, and doesn't need to wait until the final outcome

    \item This approach has an easily understood practical interpretation (will you win money betting this model) and is fundamentally Bayesian in its approach

    \item The approach works naturally in a situation where you may not view each model with equal \emph{a priori} credibility. As with Bayesian analysis in general, one is free to specify pre-game priors (i.e.\ bankrolls) for each model. For example, we could have started Alice with a \$75 bankroll and Bob with a \$25 bankroll in the opening example.

    \item Building on the previous point, the approach can be implemented in a sequential way, with the ending bankroll from a previous contest carrying over into the subsequent contest, just as one would do with an updated Bayesian prior. We saw in Section~\ref{sec:iteration} that this sequential evaluation feature results in a clear advantage in assessing model accuracy.
\end{itemize}

%----------------------------------------------------------------------
\section{Conclusion}
\label{sec:conclusion}
%----------------------------------------------------------------------

From election models to in-game sports wagering to rapidly growing prediction markets such as Polymarket and Kalshi, probabilistic forecasts that update over time are having a moment, as they say. This increasing prevalence calls for a rigorous evaluation framework with which to judge the accuracy and utility of these types of predictions---one that is tailored to their time updating nature and improves upon existing, simpler methods, such as averaging log loss or Brier scores. This paper provides a flexible and easily applied framework to evaluate such forecasts in real time, with both solid theoretical and practical underpinnings.

%----------------------------------------------------------------------
\section*{Acknowledgments}
%----------------------------------------------------------------------

The author thanks Michael Lopez for his advice and guidance on the structure and focus of this paper, particularly his suggestion to use simulations to demonstrate the utility of this approach. Thank you to Brian Burke for providing ESPN win probability data for the NFL evaluation example. The author also thanks Andrew Patton and Rajiv Sethi for their advice and encouragement.

%----------------------------------------------------------------------
\appendix
\renewcommand{\theequation}{A.\arabic{equation}}
\setcounter{equation}{0}
\section{Formula Derivations}
\label{app:derivations}
%----------------------------------------------------------------------

\subsection{The Kelly Criterion when there are existing bets}
\label{app:kelly_existing}

We can generalize the Kelly criterion to the case where a bettor has existing positions by returning to the core definition, which is that Kelly bettors seek to optimize the expected log of their long term wealth. We will denote $w$ as a bettor's \emph{win shares}, and define it as the money a bettor would stand to win if their bets were successful. When a bettor has existing win shares, $w$, the expected log of their long term wealth, $G$, can be expressed as follows:
\[
G = p \log(b + bfo + w) + (1-p) \log(b - fb)
\]

Maximizing this expression with respect to $f$ and solving for $f$:
\[
\frac{pbo}{b + fbo + w} - \frac{(1-p)b}{b - bf} = 0
\]

\[
f = p - \frac{(1-p)}{o}\left(1+\frac{w}{b}\right)
\eqno{(1)}
\]

\subsection{Determining a market clearing price for binary probabilities}
\label{app:market_clearing}

This market clearing price condition can be expressed formally as:
\begin{equation}
\sum_{i=1}^{n} f_i b_i = 0
\label{eq:app_market_clear_cond}
\end{equation}

Where $f_i$ is the fraction of bankroll bet (or fraction accepted for a bet if $f_i$ is negative), and $b_i$ is the current bankroll for the $i$th bettor. If $f_i$ is negative, this means the bettor is accepting, rather than placing a bet.

One can combine Equations~\eqref{eq:kelly_existing} and~\eqref{eq:app_market_clear_cond} and then solve for $o$, the market clearing odds that ensures each bettor is either placing or accepting a bet according to the Kelly criterion. To simplify the formulas, and to better draw the connection to Bayesian credibility, it is assumed that the combined bankroll of all $n$ participants sums to 1:
\begin{equation}
\sum_{i=1}^{n} b_i = 1
\label{eq:app_bankroll_sum}
\end{equation}

In addition, it is assumed that all market participants start off with zero win shares. Since win shares can only be exchanged on a zero-sum basis and cannot be created or destroyed, the total sum of all held win shares among market participants must always sum to zero:
\begin{equation}
\sum_{i=1}^{n} w_i = 0
\label{eq:app_winshares_sum}
\end{equation}

The market clearing odds can be derived by combining formulas~\eqref{eq:kelly_existing} and~\eqref{eq:app_market_clear_cond}:
\[
\sum_{i=1}^{n}\left( p_i b_i - \frac{(1-p_i)}{o}b_i - \frac{(1-p_i)w_i}{o} \right) = 0
\]

Solving for $o$, we find that the ``market clearing'' price for a collection of $n$ Kelly bettors with existing bets is:
\[
\text{market odds} = \frac{1-\sum p_i b_i}{\sum p_i b_i} - \frac{\sum p_i w_i}{\sum p_i b_i}
\]

Where equations~\eqref{eq:app_bankroll_sum} and~\eqref{eq:app_winshares_sum} have been used to simplify the expression.

This can be expressed even more simply in terms of the probability implied by these odds:
\[
\text{market probability} = \frac{\sum p_i b_i}{1-\sum p_i w_i}
\eqno{(2)}
\]

\subsection{Derivation of Kelly betting with multiple outcomes and existing bets}
\label{app:kelly_multi}

Following Kelly's original formulation, assume that the market odds are ``fair'', in that there is no ``track take'' or ``vigorish'' baked into the odds. This condition can be expressed as:
\[
\sum_{i=1}^{n} m_i = 1
\]

Assume now that our hypothetical bettor seeks to maximize the expected logarithm of their long term wealth, as dictated by the Kelly criterion. This can only be achieved by making and/or accepting bets at the given market odds. Making bets in this manner shifts their win shares from $w_i$ to $w_i'$, which we'll express as $w_i' = w_i + \Delta_i$. This can be treated as a constrained optimization problem, where the bettor seeks to optimize the expected logarithm of their wealth via $\Delta_i$, subject to the constraint that $\Delta_i$ must be in accordance with market odds. What this means in practice is that the bettor must fund any bets via \emph{accepting} bets on different outcomes. The market odds are, in effect, the exchange rate for swapping win shares.

This constraint on $\Delta_i$ can be expressed formally as:
\[
\sum_{i=1}^{n} \Delta_i m_i = 0
\]

To illustrate with an example, if the market probabilities for a game were 0.25 (3 to 1 odds) for the home team to win and 0.75 for the away team to win (1 to 3 odds), and you wished to increase your win shares on the home team by \$2, you would first make a \$0.50 bet on the home team at 3 to 1 odds. In order to fund that bet, you would need to accept a bet on the away team at \$0.50, which is the equivalent of selling \$0.67 win shares on the away team. \$2 $\times$ 0.25 $-$ \$0.67 $\times$ 0.75 = 0, in accordance with the formula above.

Calculating $\Delta_i$ is a constrained optimization problem, which can be handled via a Lagrange multiplier. Optimizing the expected log of one's bankroll subject to the constraint imposed by the market odds requires one to optimize the following expression with respect to $\Delta_i$ and $\lambda$:
\[
\sum_{i=1}^{n} p_i \cdot \log\left( w_i + \Delta_i\right) - \lambda \sum_{i=1}^{n}\Delta_i m_i
\]

Where $\lambda$ is the Lagrangian multiplier used to impose the constraint. Differentiating with respect to $\Delta_i$ and setting to zero:
\begin{equation}
\frac{p_i}{w_i + \Delta_i} - \lambda m_i = 0
\label{eq:app_lagrange}
\end{equation}

Solve for $\lambda$ first by summing over $i$:
\[
\sum_{i=1}^{n}p_i = \lambda \sum_{i=1}^{n} m_i w_i + \lambda \sum_{i=1}^{n} m_i \Delta_i
\]

The term on the left side of the equation sums to 1, being a sum of probabilities over all possible outcomes. And the 2nd term on the right side sums to zero, per the original constraint. Solving for the Lagrangian multiplier $\lambda$:
\[
\lambda = \frac{1}{\sum_{i=1}^{n}m_i w_i}
\]

One can now plug this into equation~\eqref{eq:app_lagrange} above and solve for $\Delta_i$:
\begin{equation}
\Delta_i = \frac{p_i}{m_i} \sum_{k=1}^{n} m_k w_k - w_i
\label{eq:delta}
\end{equation}

And since $w_i' = w_i + \Delta_i$, one can express the end state win shares for each bettor simply as:
\[
w'_i = \frac{p_i}{m_i} \sum_{i=1}^{n} m_i w_i
\eqno{(3)}
\]

\subsection{Derivation of market-clearing odds when there are multiple outcomes}
\label{app:market_multi}

The condition for market clearing odds is as follows:
\begin{equation}
\sum_{j=1}^{m} \Delta_{ij} = 0
\label{eq:app_market_clear_multi}
\end{equation}

For each outcome $i$, the net change in win shares among market participants must sum to zero. Win shares cannot be created or destroyed, only swapped between bettors in accordance with the market odds.

Combining formulas~\eqref{eq:delta} and~\eqref{eq:app_market_clear_multi}, we can now solve for the market clearing probabilities $m_i$.
\[
\sum_{j=1}^{m} \left(\frac{p_{ij}}{m_i} \sum_{k=1}^{n} m_k w_{kj} - w_{ij}\right) = 0
\]

Like win shares, the market's overall bankroll (which we have set to 1) cannot change. What this means is that for each potential outcome $i$, the sum of win shares across all bettors must sum to 1 (the money has to go somewhere):
\[
\sum_{j=1}^{m}w_{ij} = 1
\]

Combining this with our previous formula, we obtain the following condition for the market clearing probabilities $m_i$:
\[
\sum_{j=1}^{m}\sum_{k=1}^{n}p_{ij}w_{kj}m_k = m_i
\]

If we consider $p_{ij}$ and $w_{kj}$ as two $n \times m$ matrices $\bm{p}$ and $\bm{w}$, and the market probabilities $m_i$ as a vector $\bm{m}$, we can rewrite this equation in matrix form as:
\[
(\bm{p}\bm{w}^T) \bm{m} = \bm{m}
\eqno{(4)}
\]

\subsection{Connection to the PageRank Algorithm}
\label{app:pagerank}

The process for solving for the credibility vector has a close mathematical connection to the PageRank algorithm \citep{brin1998}, developed by Larry Page of Google.

The PageRank algorithm is based on two key principles:

\begin{enumerate}
    \item A web page's value, or rank, can be determined by the number of web pages that link to it
    \item Links from high value web pages should count more when determining a web page's rank
\end{enumerate}

So, PageRank defines a web page as high value if it receives inbound links from high value web pages. This definition, on its surface, seems circular. But the PageRank algorithm is a way to bootstrap one's way into a self-consistent ranking of web pages that resolves this circularity. This involves creating a ranking matrix $A$, in which the $ij$th element tells you what percentage of outbound links from web page $j$ go to web page $i$. This is web page $j$'s ``vote'' for web page $i$. Solving for the PageRank vector, $R$, amounts to finding the dominant eigenvector of the ranking matrix $A$.

The math behind the PageRank algorithm is more complex than the math used in this paper, owing to the realities of web pages and how they link to each other. Page needed to account for closed loops, and had special adjustments for ``dangling links'' in order to make the math of the PageRank algorithm workable. These mathematical difficulties do not occur in our self-evaluating market matrix $\bm{S}$.

Similar to PageRank, the procedure in Section~\ref{sec:multinomial} for evaluating the credibility of each model relies on the following:

\begin{enumerate}
    \item The credibility (or likelihood of being correct) of each model is determined by how each modeler evaluates that model's existing portfolio of bets
    \item Models with higher credibility count for more in the evaluation of each model
\end{enumerate}

In the same way that PageRank exploited the fact that web pages are both a recipient of inbound links and a generator of outbound links, our procedure for solving for the credibility vector $\bm{c}$ leverages the fact that each model has a portfolio of bets that can be evaluated, as well as its own set of probabilities with which it can evaluate all the other models (including its own).

Larry Page used the concept of a ``random surfer'' to explain the meaning of PageRank. If you have an internet user that randomly clicks on outbound links, going from web page to web page, the PageRank vector tells you the amount of time one would expect to spend on each site as a ``random surfer''. More technically, it is the steady state probability distribution for the Markov process described by the random surfer's browsing habits.

We can extend the random surfer analogy to our self-evaluating market. Rather than random surfers, we have random bettors. Pick a bettor in our market at random and have them evaluate the expected value of each bettor's portfolio according to their own probability estimates (this just corresponds to a column in our matrix $\bm{S}$). We then randomly jump to a new bettor, where the probability of jumping from bettor A to bettor B is just bettor A's subjective evaluation of bettor B's portfolio. Repeat this process indefinitely, and the proportion of time one jumps to a specific bettor A corresponds to that bettor's credibility, as well as the steady state distribution of the Markov process.

%----------------------------------------------------------------------
\section{Example Applications of the Methodology}
\label{app:examples}
%----------------------------------------------------------------------

\subsection{NFL In-Game Win Probability}
\label{app:nfl}

We will first illustrate the methodology with four examples from the NFL. There are several models that forecast the probability of an NFL team winning a game based on game situation. For this illustration we will use the win probability model developed by ESPN Analytics and the Open Source Football (OSF) win probability model that can be accessed through the R package nflfastR. Table~\ref{tab:nfl_summary} summarizes the results.

\begin{table}[H]
\centering
\caption{Model performance for Open Source Football (OSF) and ESPN, summarized for four NFL games. The ``winning'' model score is underlined.}
\label{tab:nfl_summary}
\small
\begin{tabular}{l cc cc cc}
\toprule
& \multicolumn{2}{c}{\textbf{log loss}} & \multicolumn{2}{c}{\textbf{brier score}} & \multicolumn{2}{c}{\textbf{kelly credibility}} \\
\cmidrule(lr){2-3}\cmidrule(lr){4-5}\cmidrule(lr){6-7}
\textbf{game} & OSF & ESPN & OSF & ESPN & OSF & ESPN \\
\midrule
2/4/18 -- PHI @ NE & 0.805 & \underline{0.580} & 0.192 & \underline{0.123} & 46.2\% & \underline{53.8\%} \\
1/29/23 -- CIN @ KC & 0.619 & \underline{0.524} & 0.128 & \underline{0.100} & 45.4\% & \underline{54.6\%} \\
12/10/23 -- LA @ BAL & \underline{0.775} & 0.873 & \underline{0.176} & 0.206 & \underline{65.4\%} & 34.6\% \\
12/18/23 -- PHI @ SEA & 2.354 & \underline{2.111} & 0.633 & \underline{0.576} & \underline{60.2\%} & 39.8\% \\
\midrule
\textbf{Average} & \textbf{1.138} & \textbf{\underline{1.022}} & \textbf{0.282} & \textbf{\underline{0.251}} & \textbf{\underline{54.3\%}} & \textbf{45.7\%} \\
\bottomrule
\end{tabular}
\end{table}

\begin{figure}[H]
\centering
\includegraphics[width=0.85\textwidth]{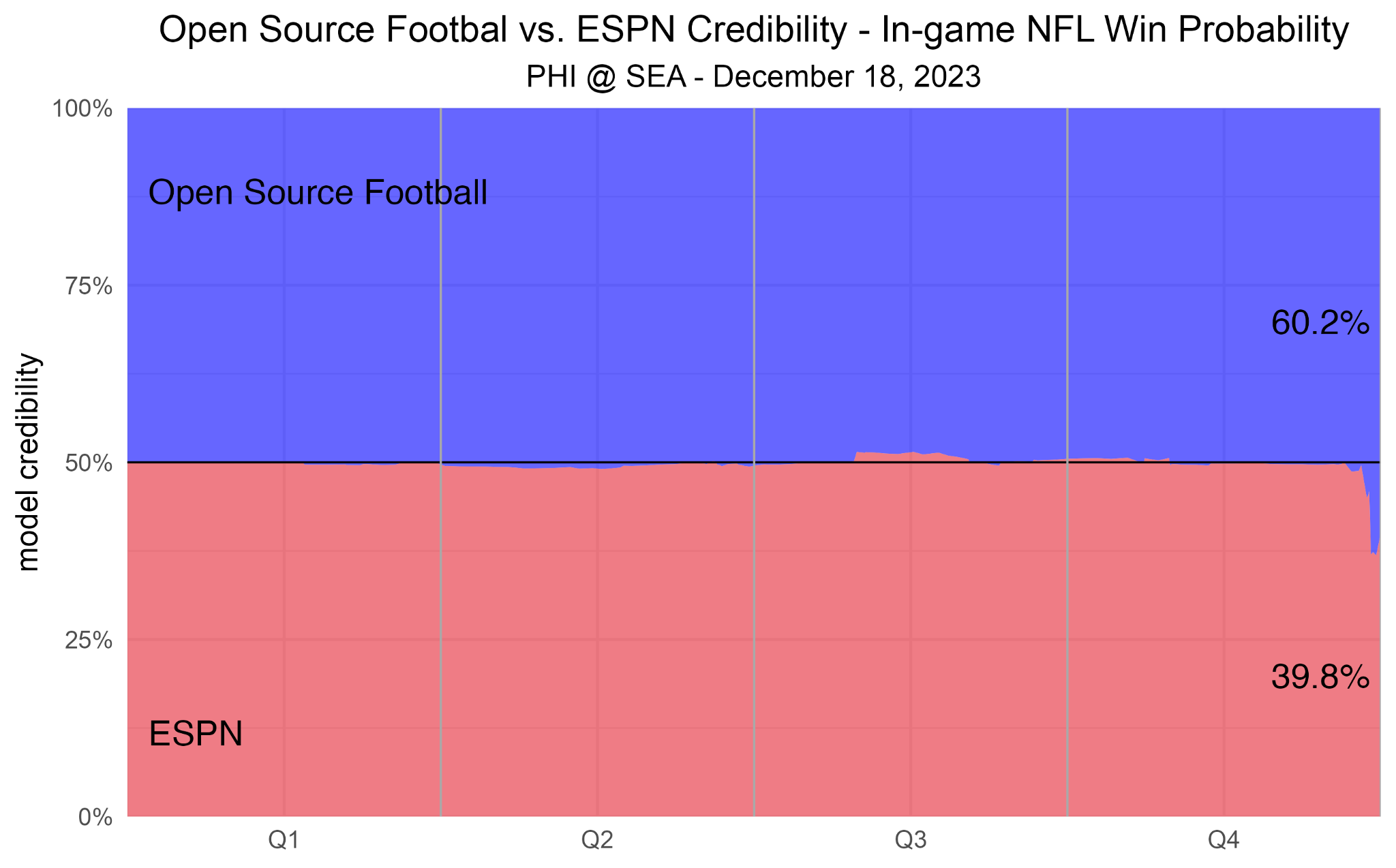}
\caption{Implied model credibility for Open Source Football and ESPN, evaluated in-game}
\label{fig:phi_sea_cred}
\end{figure}

The PHI @ SEA game is an instructive example, as it's a game where the more traditional average log loss and Brier score methods yield a different result than the Kelly-based approach. From Figure~\ref{fig:phi_sea_wp} below, one can see that the ESPN model had Seattle with the higher win probability for most of the game, but that flips midway through the 4th quarter. From Figure~\ref{fig:phi_sea_cred}, we see that the implied credibility of the two models did not move much for most of the game, until the final minute.

\begin{figure}[H]
\centering
\includegraphics[width=0.85\textwidth]{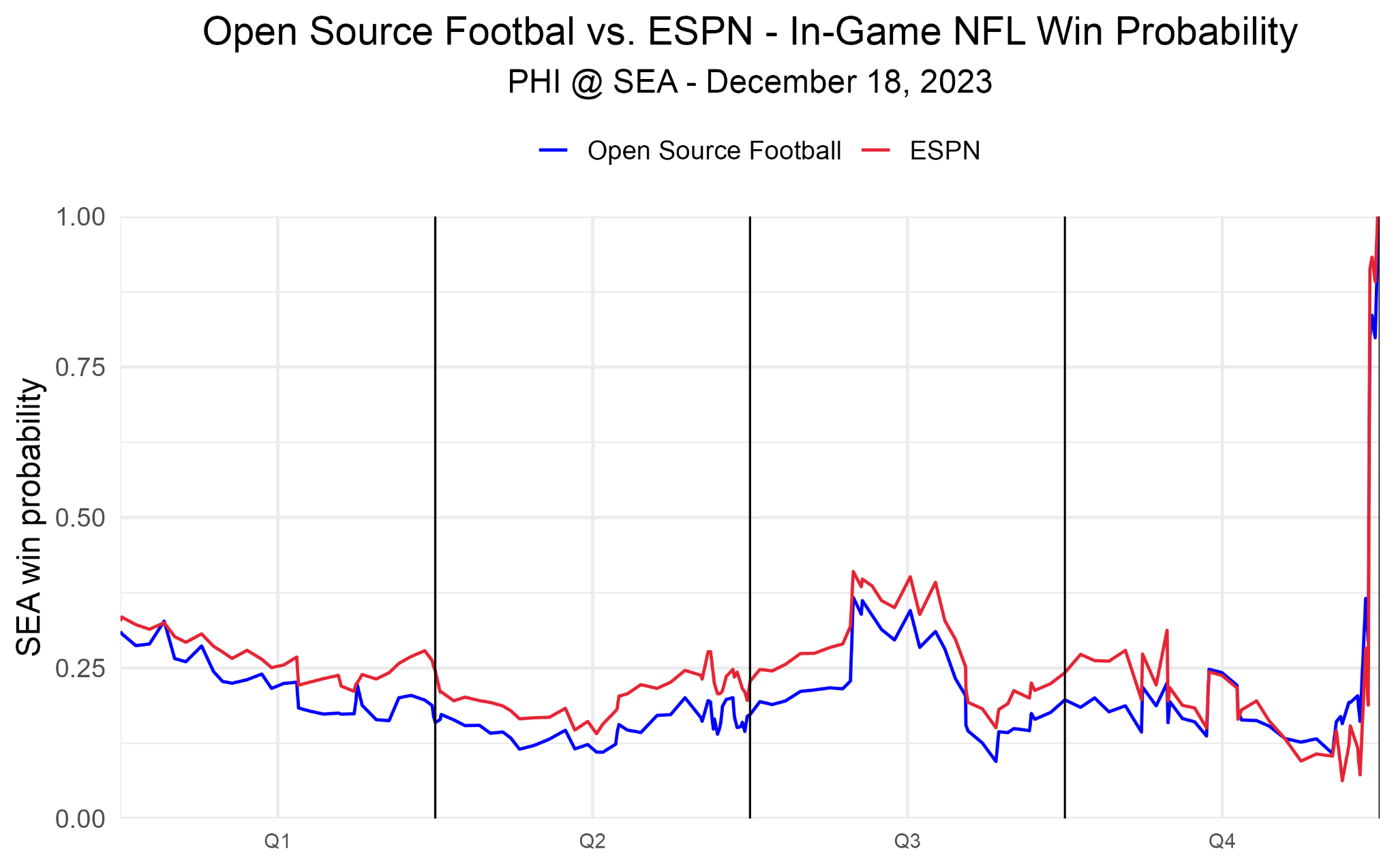}
\caption{Comparison of in-game win probabilities for Open Source Football and ESPN for the Seahawks--Eagles game on December 18, 2023}
\label{fig:phi_sea_wp}
\end{figure}

Table~\ref{tab:phi_sea_state} shows game state and corresponding win probability estimates for both models for the final minute. Table~\ref{tab:phi_sea_market} shows how those probability estimates are converted into market odds, bets, and real-time model credibilities. The biggest drop in credibility for the ESPN model occurs when the market win probability flips from 24\% to 85\% (the Seahawks, trailing by 4, scored a touchdown on 3rd and 10 from the Eagles' 29 yard line, with 0:28 left to go). ESPN's credibility takes a hit because of the \emph{relative} likelihood it gave the Seahawks (18.8\%), compared to Open Source Football (29.4\%).

\begin{table}[H]
\centering
\caption{Game state and corresponding win probabilities from the Open Source Football (OSF) and ESPN models for the final minute of the 12/18/23 PHI @ SEA game}
\label{tab:phi_sea_state}
\small
\begin{tabular}{cclcccccc}
\toprule
& & & & & & \multicolumn{2}{c}{\textbf{score}} & \textbf{win prob} \\
\cmidrule(lr){7-8}
\textbf{q} & \textbf{clock} & \textbf{pos} & \textbf{yl} & \textbf{down} & \textbf{togo} & PHI & SEA & OSF \quad ESPN \\
\midrule
4 & 1:00 & SEA & 63 & 2 & 10 & 17 & 13 & 18.3\% \quad 9.5\% \\
4 & 0:56 & SEA & 63 & 3 & 10 & 17 & 13 & 16.1\% \quad 7.3\% \\
4 & 0:40 & SEA & 29 & 1 & 10 & 17 & 13 & 36.6\% \quad 28.3\% \\
4 & 0:36 & SEA & 29 & 2 & 10 & 17 & 13 & 33.7\% \quad 22.6\% \\
4 & 0:33 & SEA & 29 & 3 & 10 & 17 & 13 & 29.4\% \quad 18.8\% \\
4 & 0:28 & PHI & 75 & 1 & 10 & 17 & 20 & 80.0\% \quad 91.3\% \\
\bottomrule
\end{tabular}
\end{table}

\begin{table}[H]
\centering
\caption{Resulting market odds, bets, and model credibility results for the Open Source Football (OSF) and ESPN win probability models, for the final minute of the 12/18/23 PHI @ SEA game}
\label{tab:phi_sea_market}
\small
\begin{tabular}{cc cc c cc cc cc}
\toprule
& & \multicolumn{2}{c}{\textbf{win prob}} & & \multicolumn{2}{c}{\textbf{bankroll}} & \multicolumn{2}{c}{\textbf{win shares}} & \multicolumn{2}{c}{\textbf{credibility}} \\
\cmidrule(lr){3-4}\cmidrule(lr){6-7}\cmidrule(lr){8-9}\cmidrule(lr){10-11}
\textbf{q} & \textbf{clock} & OSF & ESPN & Market & OSF & ESPN & OSF & ESPN & OSF & ESPN \\
\midrule
4 & 1:00 & 18.3\% & 9.5\% & 13.9\% & 0.486 & 0.514 & 0.161 & $-$0.161 & 0.509 & 0.491 \\
4 & 0:56 & 16.1\% & 7.3\% & 11.6\% & 0.482 & 0.518 & 0.189 & $-$0.189 & 0.504 & 0.496 \\
4 & 0:40 & 36.6\% & 28.3\% & 32.5\% & 0.479 & 0.521 & 0.220 & $-$0.220 & 0.550 & 0.450 \\
4 & 0:36 & 33.7\% & 22.6\% & 28.5\% & 0.518 & 0.482 & 0.101 & $-$0.101 & 0.546 & 0.454 \\
4 & 0:33 & 29.4\% & 18.8\% & 24.4\% & 0.506 & 0.494 & 0.140 & $-$0.140 & 0.540 & 0.460 \\
4 & 0:28 & 80.0\% & 91.3\% & 84.9\% & 0.505 & 0.495 & 0.148 & $-$0.148 & 0.630 & 0.370 \\
\bottomrule
\end{tabular}
\end{table}

\subsection{Major League Baseball Division Win Probabilities}
\label{app:mlb}

The second example comes from Major League Baseball and will allow us to evaluate modeling of multinomial probabilities. For the 2022 MLB season, both FiveThirtyEight \citep{538mlb} and FanGraphs \citep{fangraphs} published forecasted probabilities of each team winning their respective division throughout the season. FiveThirtyEight archived these probabilities on a weekly basis. These were then merged with the corresponding FanGraphs forecast. Each division was treated as a separate betting contest, with the results summarized in Table~\ref{tab:mlb_summary} below.

\begin{table}[H]
\centering
\caption{Performance of the FiveThirtyEight and FanGraphs division winner forecasts (winning metric is underlined)}
\label{tab:mlb_summary}
\small
\begin{tabular}{l cc cc cc}
\toprule
& \multicolumn{2}{c}{\textbf{log loss}} & \multicolumn{2}{c}{\textbf{brier score}} & \multicolumn{2}{c}{\textbf{kelly credibility}} \\
\cmidrule(lr){2-3}\cmidrule(lr){4-5}\cmidrule(lr){6-7}
\textbf{division} & 538 & FanGraphs & 538 & FanGraphs & 538 & FanGraphs \\
\midrule
AL East & \underline{0.065} & 0.071 & \underline{0.028} & 0.033 & \underline{51.4\%} & 48.6\% \\
AL Central & \underline{0.276} & 0.310 & \underline{0.155} & 0.171 & \underline{55.8\%} & 44.2\% \\
AL West & 0.047 & \underline{0.029} & 0.018 & \underline{0.010} & 45.0\% & \underline{55.0\%} \\
NL East & \underline{0.294} & 0.311 & \underline{0.176} & 0.189 & 45.9\% & \underline{54.1\%} \\
NL Central & \underline{0.217} & 0.255 & \underline{0.133} & 0.159 & \underline{52.8\%} & 47.2\% \\
NL West & \underline{0.040} & 0.062 & \underline{0.012} & 0.025 & \underline{54.6\%} & 45.4\% \\
\midrule
\textbf{Average} & \textbf{\underline{0.156}} & \textbf{0.173} & \textbf{\underline{0.087}} & \textbf{0.098} & \textbf{\underline{50.9\%}} & \textbf{49.1\%} \\
\bottomrule
\end{tabular}
\end{table}

\begin{figure}[H]
\centering
\includegraphics[width=0.95\textwidth]{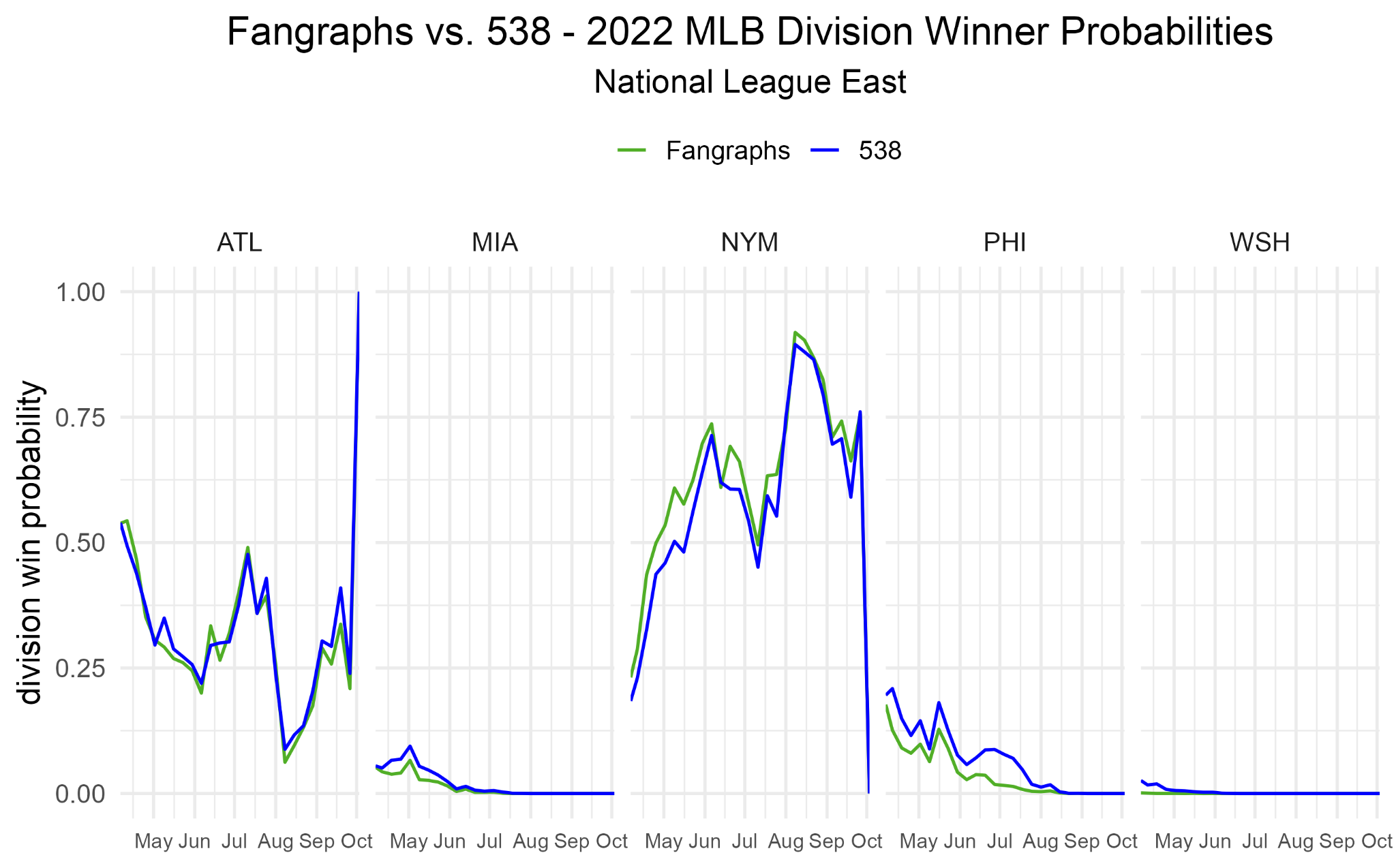}
\caption{Weekly division win probabilities for each team, shown separately for FanGraphs and FiveThirtyEight}
\label{fig:nleast_wp}
\end{figure}

\begin{figure}[H]
\centering
\includegraphics[width=0.95\textwidth]{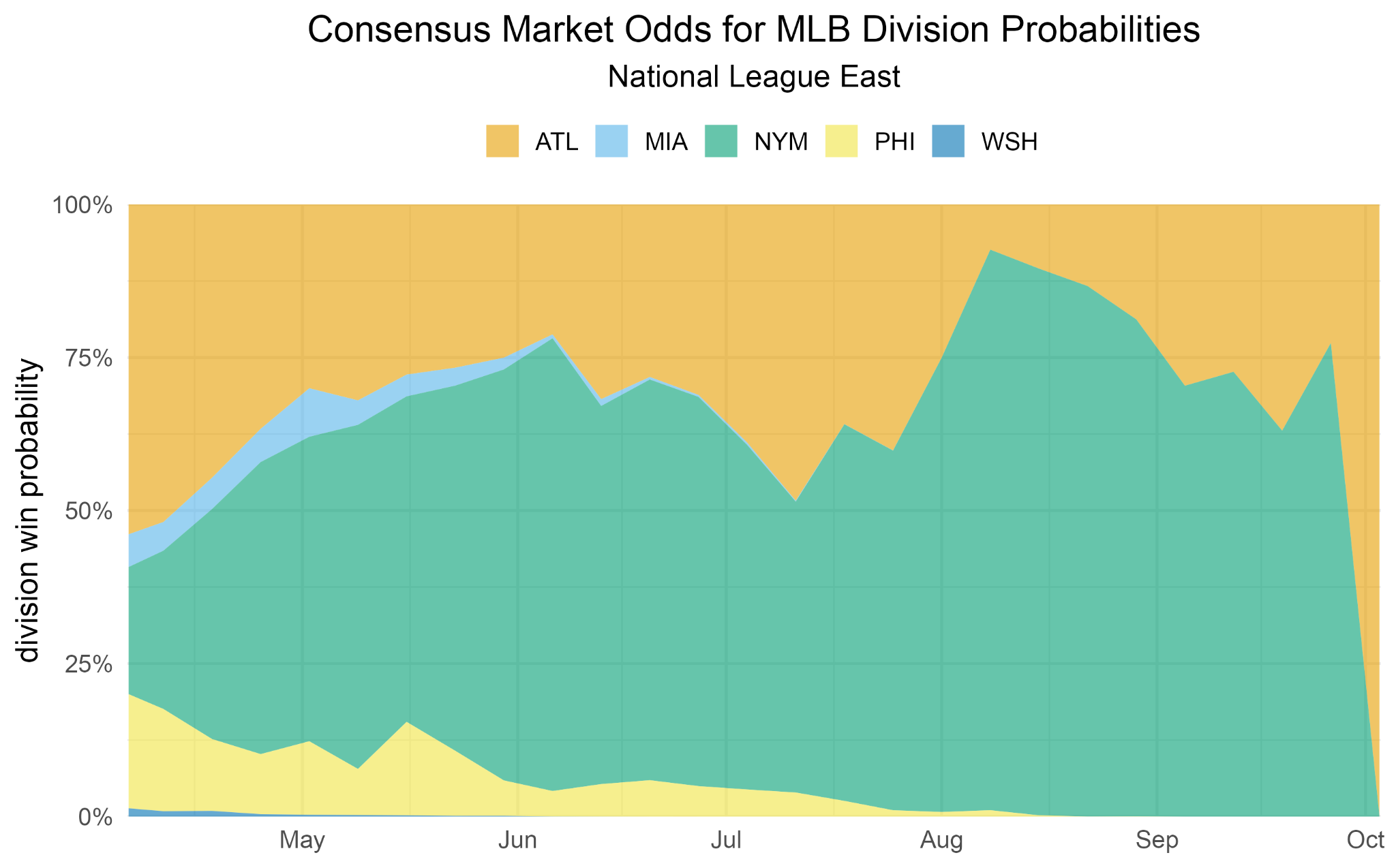}
\caption{Market consensus division win probabilities by week for the 2022 NL East Division}
\label{fig:nleast_market}
\end{figure}

\begin{figure}[H]
\centering
\includegraphics[width=0.95\textwidth]{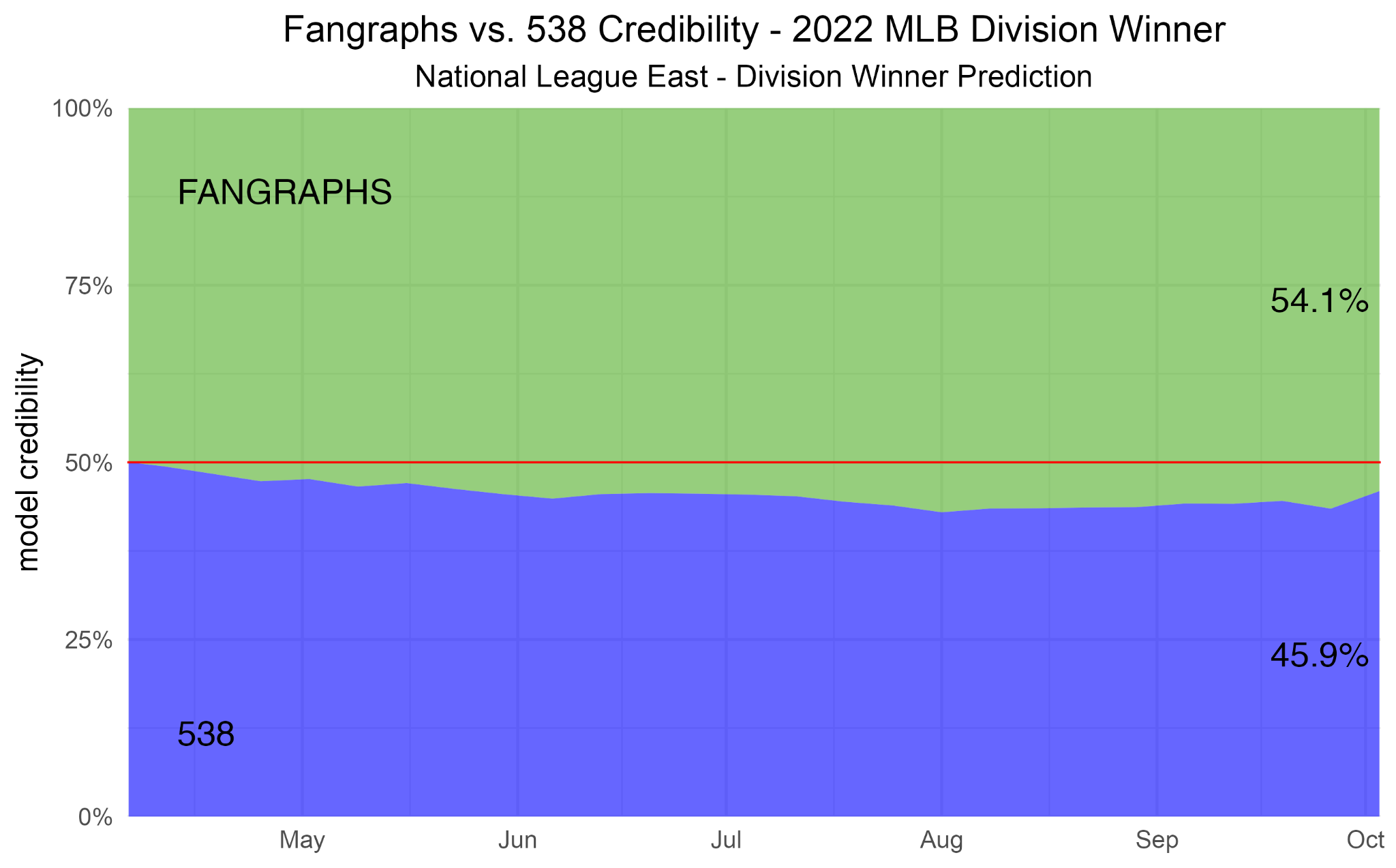}
\caption{Weekly model credibility for FanGraphs and FiveThirtyEight, based on the 2022 NL East division race}
\label{fig:nleast_cred}
\end{figure}

Figures~\ref{fig:nleast_wp}--\ref{fig:nleast_cred} illustrate the methodology for the NL East division race.

This is another example where the Kelly results are at odds with the traditional log loss and Brier score results. The FiveThirtyEight model loses credibility by August because of the growth in the Mets' win probability and the decline in the Phillies' win probability. In both cases, FanGraphs modeling was more prescient than FiveThirtyEight's as seen in Figure~\ref{fig:nleast_wp}.

\subsection{NBA In-Game Win Probability}
\label{app:nba}

In this example, we will show how the approach can be applied sequentially throughout a series of games, where the ending bankroll/credibility at one game serves as the starting bankroll for the next game. The two probability models to be compared are FiveThirtyEight's NBA In-Game win probability model \citep{538nba} and the author's NBA In-Game win probability published at inpredictable.com \citep{inpredictable}.

The models will bet against each other for the entirety of the 2023 NBA playoffs. Games are ordered sequentially so that the ending bankroll/credibility numbers from one game are used as the starting point for the next game. This allows for a more substantive and Bayesian-focused test of the accuracy of the models.

\begin{figure}[H]
\centering
\includegraphics[width=0.95\textwidth]{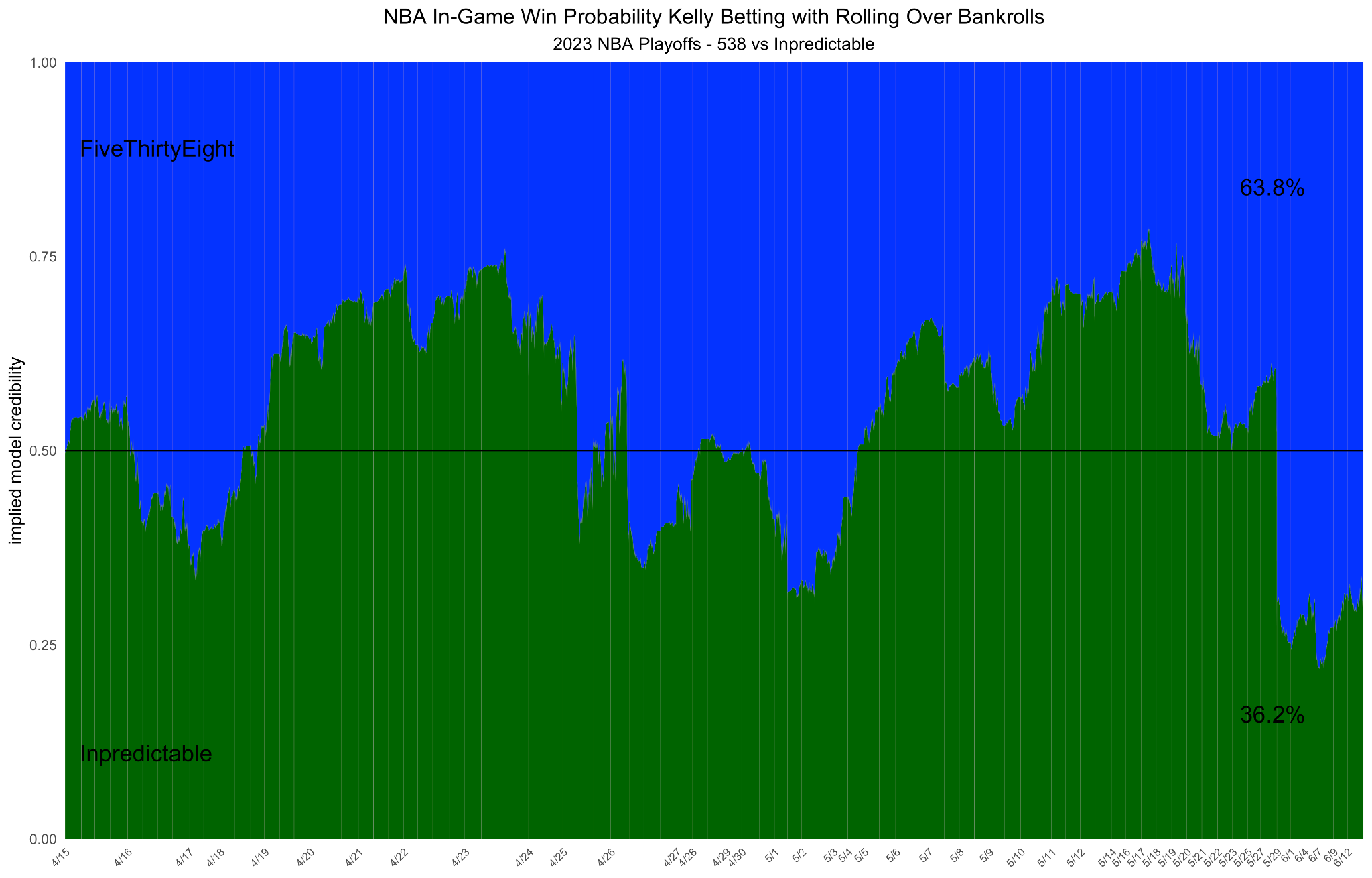}
\caption{In-game credibility for the FiveThirtyEight and Inpredictable in-game probability models, calculated sequentially over the course of the 2023 NBA playoffs}
\label{fig:nba_cred}
\end{figure}

Figure~\ref{fig:nba_cred} illustrates the play-by-play and game-by-game evolution of the credibility of the Five\-Thirty\-Eight and the inpredictable models. Each lightly shaded line represents the start of a new game. Both models begin at 50\% credibility. After 84 games and 31,000+ plays, the FiveThirtyEight model ends up with net credibility gain of +13.8\%.

Towards the end of the playoffs, note the sharp shift in credibility from Inpredictable to FiveThirtyEight. This shift can be traced to a single play at the conclusion of Game 6 of the Eastern Conference Finals. With 3 seconds left, the Celtics trailed the Heat by 1. Off the inbounds pass, Marcus Smart's three pointer was off the mark, but Derrick White was able to get the rebound plus the tip in to win the game at the buzzer for Boston.

The play by play data records this as an offensive rebound for the Celtics, followed by the game winning shot. At the point of the offensive rebound (down 1, with 1 second left), but \emph{prior} to the game winner, FiveThirtyEight gave the Celtics a 19\% chance of victory, while Inpredictable gave the Celtics just a 5\% chance. Because the Celtics won, the net effect was to shift FiveThirtyEight's credibility from 47\% to 69\%. While this may seem like a large shift based on a single play, this does serve to highlight the Bayesian nature of this approach.

To use an example more typically employed in Bayesian reasoning, instead of competing probability models, imagine if one had two bags of white and black balls, where the 1st bag had 95 white balls and 5 black balls, and the 2nd bag had 81 white balls and 19 black balls. One of the bags is chosen at random and a ball is selected from the bag. Prior to knowing the color of the ball, the probability of the 2nd bag being the chosen bag is 50\%. But if the selected ball is black, the posterior probability of the 2nd bag being chosen jumps to 79\%. Similar to Bayesian updating, an improbable outcome shifts model bankrolls based on the assessed relative likelihood of the observed result.

%----------------------------------------------------------------------
\section{Additional Simulation Exhibits}
\label{app:sim_exhibits}
%----------------------------------------------------------------------

The charts below are supplements to the analysis summarized in Section~\ref{sec:iteration_sim}, in which a simulated betting contest is iterated over several trials.

Each square in the chart below denotes which model evaluation approach was most accurate in identifying the correct model (K=Kelly approach, L=Average log loss, B= Average Brier score). If more than one letter is displayed in a square, that means that the model evaluation approaches tied in terms of accuracy. Green squares indicate where the Kelly approach was most accurate, yellow squares indicate where it tied for highest accuracy, and red squares indicate when either the Log Loss or Brier Score approach was more accurate.

\begin{figure}[H]
\centering
\includegraphics[width=0.95\textwidth]{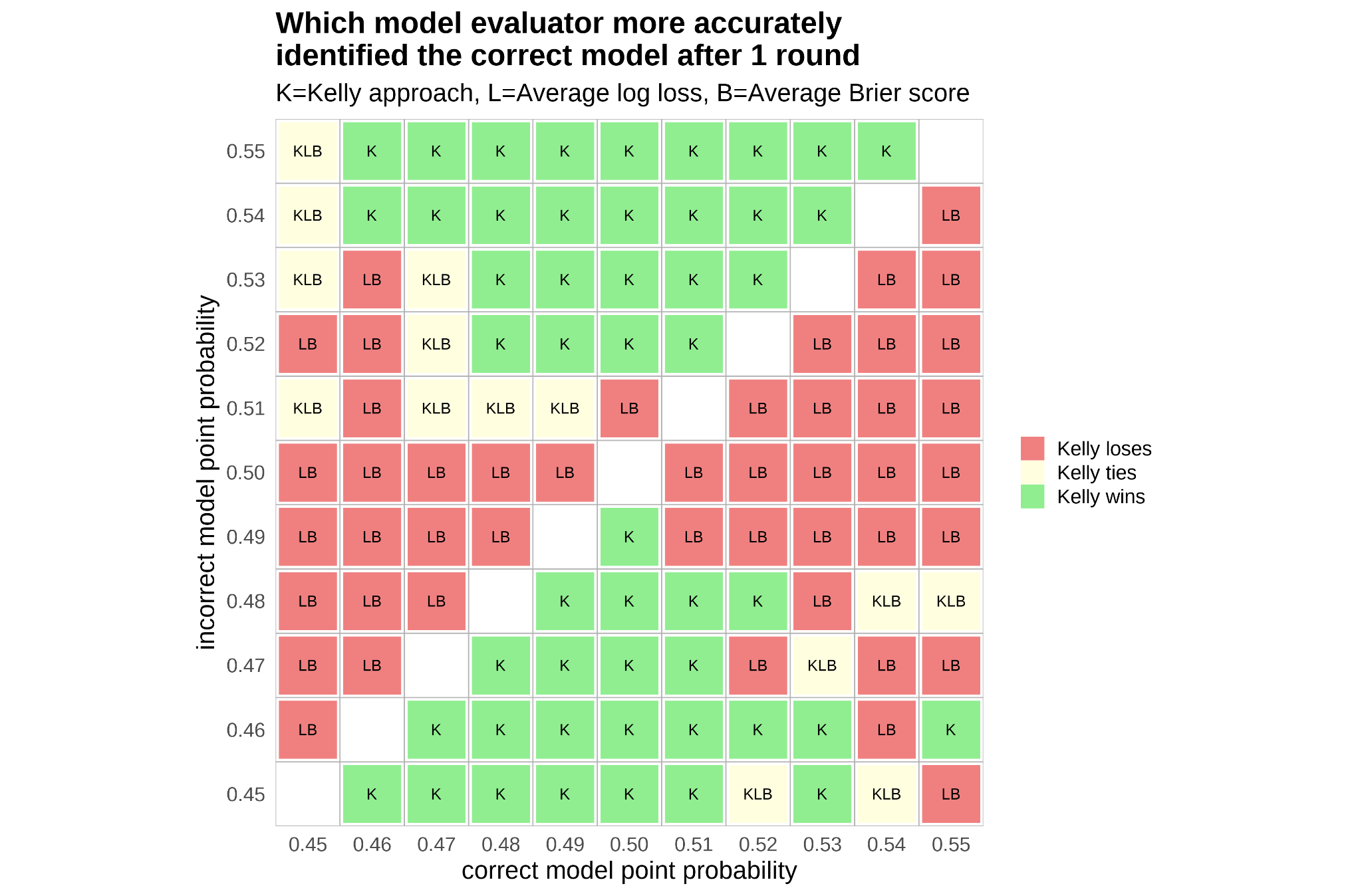}
\caption{Model evaluation accuracy grid after 1 game}
\label{fig:sim_1game}
\end{figure}

\begin{figure}[H]
\centering
\includegraphics[width=0.95\textwidth]{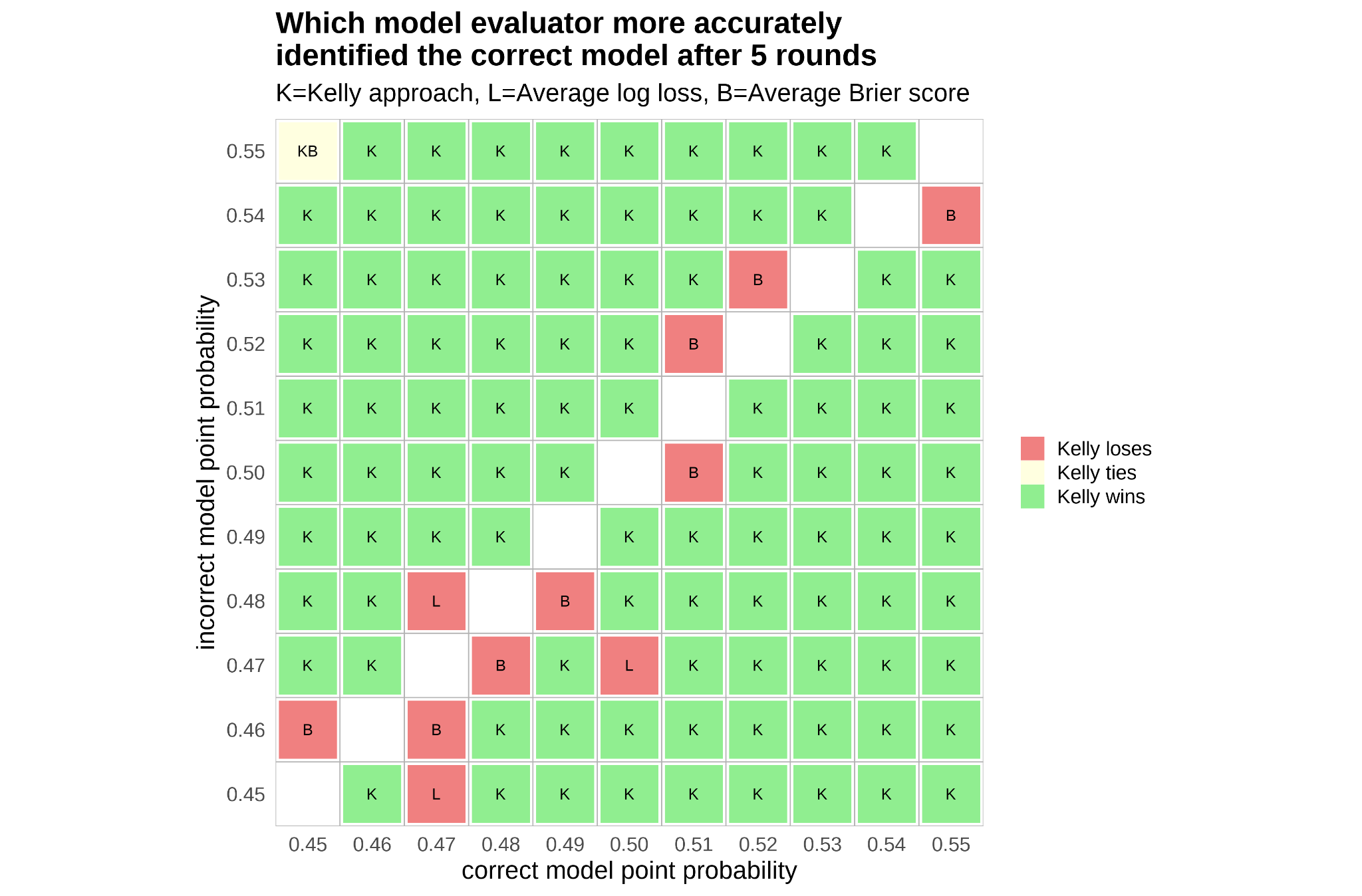}
\caption{Model evaluation accuracy grid after 5 games}
\label{fig:sim_5games}
\end{figure}

\begin{figure}[H]
\centering
\includegraphics[width=0.95\textwidth]{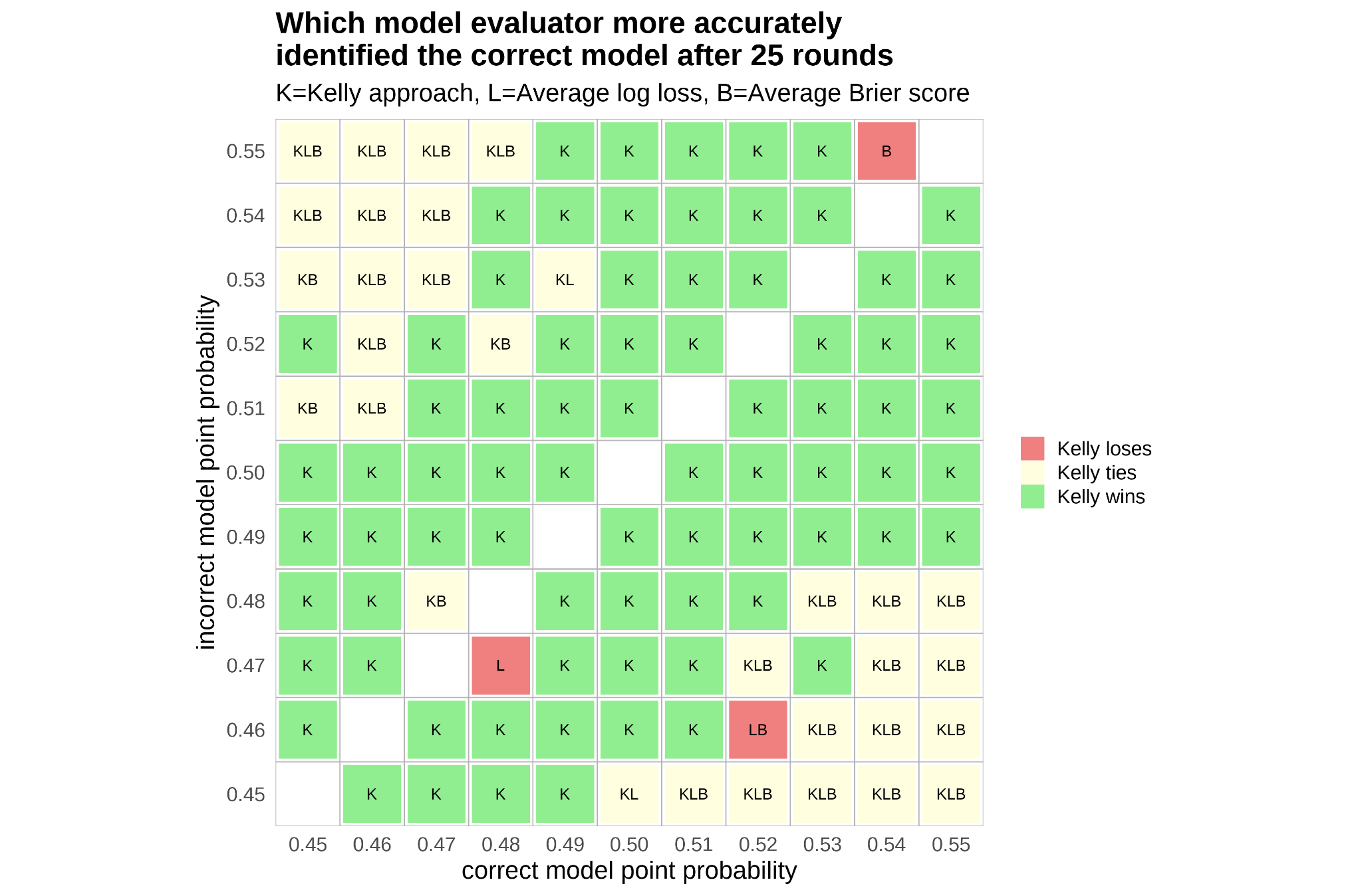}
\caption{Model evaluation accuracy grid after 25 games}
\label{fig:sim_25games}
\end{figure}

\begin{figure}[H]
\centering
\includegraphics[width=0.95\textwidth]{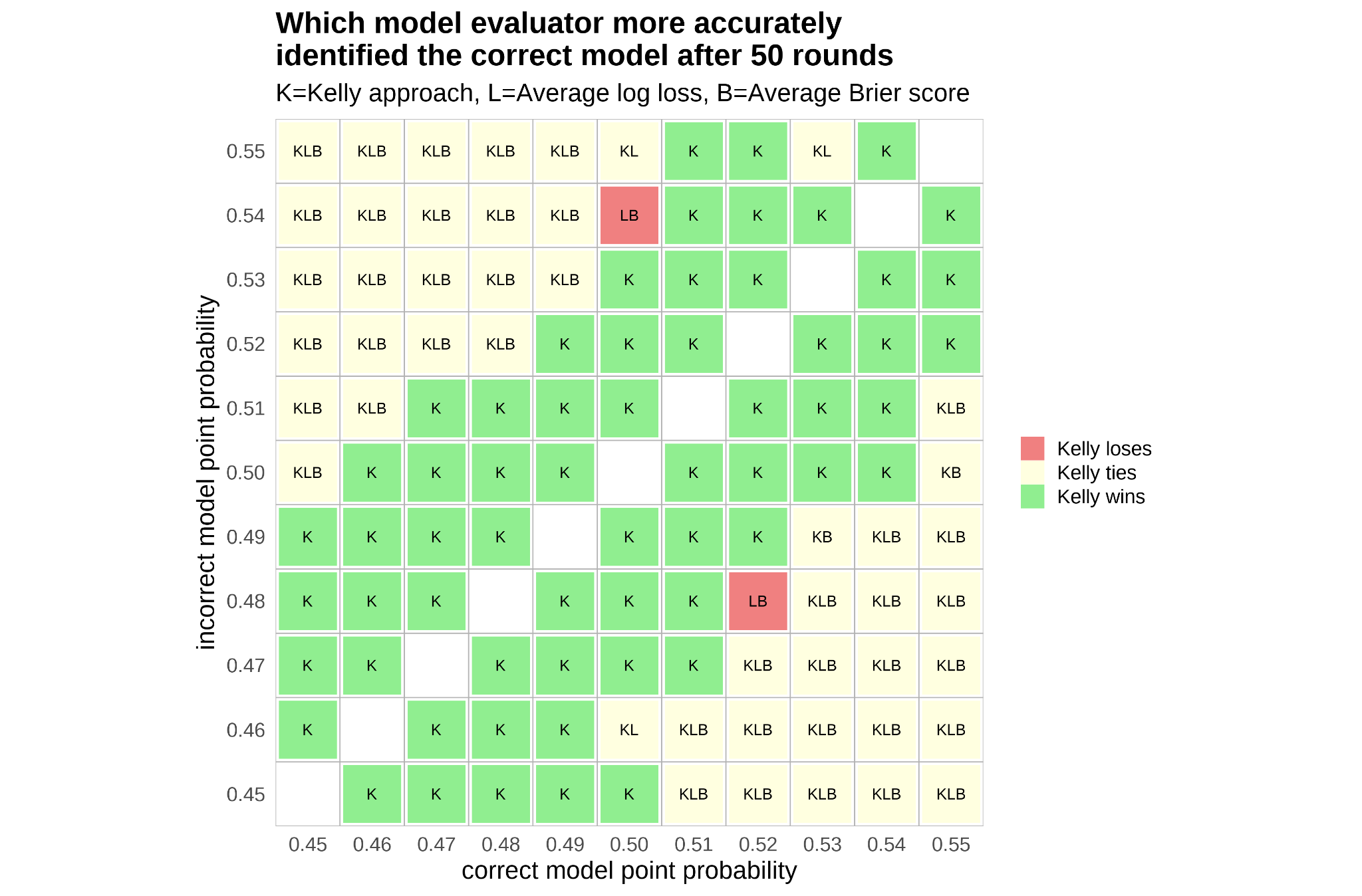}
\caption{Model evaluation accuracy grid after 50 games}
\label{fig:sim_50games}
\end{figure}

%----------------------------------------------------------------------
\bibliographystyle{plainnat}
\bibliography{references}
%----------------------------------------------------------------------

\end{document}